\documentclass[twocolumn]{aastex63}
\usepackage[T1]{fontenc}
\usepackage{ae,aecompl}

\usepackage{multirow}

\received{February 16, 2021}
\revised{May 12, 2021}
\accepted{May 12, 2021}

\submitjournal{ApJ}

\shorttitle{PMN~J0909+0354}
\shortauthors{Perger et al.}

\begin{document}

\correspondingauthor{Krisztina Perger}
\email{perger.krisztina@csfk.org}

\author[0000-0002-6044-6069]{Krisztina Perger}
\affiliation{Konkoly Observatory, ELKH Research Centre for Astronomy and Earth Sciences,
 Konkoly Thege Mikl\'os \'ut 15-17, H-1121 Budapest, Hungary}
\affiliation{Department of Astronomy, E\"otv\"os Lor\'and University,
P\'azm\'any P\'eter s\'et\'any 1/A,
H-1117 Budapest, Hungary}

\author[0000-0003-3079-1889]{S\'andor Frey}
\affiliation{Konkoly Observatory, ELKH Research Centre for Astronomy and Earth Sciences, Konkoly Thege Mikl\'os \'ut 15-17, H-1121 Budapest, Hungary}
\affiliation{Institute of Physics, ELTE E\"otv\"os Lor\'and University,
P\'azm\'any P\'eter s\'et\'any 1/A,
H-1117 Budapest, Hungary}

\author[0000-0001-8252-4753]{Daniel A. Schwartz}
\affiliation{Center for Astrophysics, Harvard \& Smithsonian, Cambridge, MA 02138, USA}

\author[0000-0003-1020-1597]{Krisztina \'E. Gab\'anyi}
\affiliation{Department of Astronomy, E\"otv\"os Lor\'and University, 
P\'azm\'any P\'eter s\'et\'any 1/A,
H-1117 Budapest, Hungary}
\affiliation{MTA-ELTE Extragalactic Astrophysics Research Group, E\"otv\"os Lor\'and University, P\'azm\'any P\'eter s\'et\'any 1/A,
 H-1117 Budapest, Hungary}
\affiliation{Konkoly Observatory, ELKH Research Centre for Astronomy and Earth Sciences,
 Konkoly Thege Mikl\'os \'ut 15-17, H-1121 Budapest, Hungary}

\author[0000-0002-0694-2459]{Leonid I. Gurvits}
\affiliation{Joint Institute for VLBI ERIC, Oude Hoogeveensedijk 4, 7991 PD Dwingeloo, The Netherlands}
\affiliation{Department of Astrodynamics and Space Missions, Delft University of Technology, Kluyverweg 1, 2629 HS Delft, The Netherlands}

\author[0000-0002-5195-335X]{Zsolt Paragi}
\affiliation{Joint Institute for VLBI ERIC, Oude Hoogeveensedijk 4, 7991 PD Dwingeloo, The Netherlands}

\title{Multi-scale Radio and X-ray Structure of the High-redshift Quasar PMN~J0909+0354}

\begin{abstract}
The high-redshift quasar PMN~J0909+0354 ($z=3.288$) is known to have a pc-scale compact jet structure, based on global 5-GHz very long baseline interferometry (VLBI) observations performed in 1992. Its kpc-scale structure was studied with the  Karl G. Jansky Very Large Array (VLA) in the radio and the \textit{Chandra} space telescope in X-rays. Apart from the north-northwestern jet component seen in both the VLA and \textit{Chandra} images at $2\farcs3$ separation from the core, there is another X-ray feature at $6\farcs48$ in the northeastern (NE) direction. To uncover more details and possibly structural changes in the inner jet, we conducted new observations at 5~GHz using the European VLBI Network (EVN) in 2019. These data confirm the northward direction of the one-sided inner jet already suspected from the 1992 observations. A compact core and multiple jet components were identified that can be traced up to $\sim0.25$~kpc projected distance towards the north, while the structure becomes more and more diffuse. A comparison with arcsec-resolution imaging with the VLA shows that the radio jet bends by $\sim30\degr$ between the two scales. The direction of the pc-scale jet as well as the faint optical counterpart found for the newly-detected X-ray point source (NE) favors the nature of the latter as a background or foreground object in the field of view. However, the extended ($\sim160$~kpc) emission around the positions of the quasar core and NE detected by the \textit{Wide-field Infrared Survey Explorer (WISE)} in the mid-infrared might suggest physical interaction of the two objects.
\end{abstract}


\section{Introduction}

Quasars, powerful active galactic nuclei (AGN) fuelled by accretion onto supermassive black holes populate the observable Universe up to redshift $z\approx7.6$ \citep{2021ApJ...907L...1W}. Even for a redshift $z \approx 3$ that was considered `very high' for quasars known two decades ago, the corresponding age of the Universe is only about 2 billion years. Studying low- and high-redshift quasars thus provides information on the evolution of this class of objects \citep[e.g.][]{1990MNRAS.247...19D,2017A&A...602A...3D} and may also help us refining cosmological models \citep[e.g.][]{1999A&A...342..378G,2017A&A...602A..79L}.

Radiation from jetted AGN in the radio is caused by synchrotron emission of relativistic charged particles, while the role of the dominant processes in X-rays is still under debate \citep[e.g.][]{2017ApJ...849...95B,2017ApJ...846..119H,2017MNRAS.466.4299L}. X-ray emission of AGN jets might originate from inverse-Compton (IC) scattering of electrons on the cosmic microwave background (CMB), boosting the CMB energy density proportionally to the square of the bulk Lorentz factor ($\Gamma^{2}$) of the relativistic jet. This IC/CMB model can explain the morphology of one-sided X-ray jets enhanced by $\Gamma \sim 10$ with a structure extending to hundreds of kpc from the galactic nucleus into the inter-galactic space. The surface brightness of radio synchrotron emission scales down with increasing redshift by $(1+z)^{-4}$, limiting the observable population in the early Universe. In case of the IC/CMB radiation, surface brightness decreases likewise, but it is balanced out by the rising energy density of CMB photons by $(1+z)^{4}$, potentially turning X-ray emission into a redshift-independent tracer of AGN jets \citep{2002ApJ...569L..23S}. 

So far, less than twenty $z>3$ radio quasars have been imaged with the \textit{Chandra X-ray Observatory} to search for kpc-scale X-ray jets. There are two clear cases when these extend beyond the known radio jet \citep{2019AN....340...30S,2020ApJ...904...57S}. Studying sources with detectable emission in both radio and X-ray bands, physical conditions derived from the observations can be compared. Applying high-resolution very long baseline interferometry (VLBI) imaging of pc-scale radio jets at multiple epochs, apparent jet component proper motions and core brightness temperatures can be measured, and physical conditions (viewing angle, bulk Lorentz factor) of high-redshift AGN jets can be determined \citep[e.g.][]{2015MNRAS.446.2921F,2018MNRAS.477.1065P,2020SciBu..65..525Z,2020NatCo..11..143A}. Currently this sample is very limited at the highest redshifts. Therefore multi-epoch VLBI imaging of another jetted object is of particular interest.    

The high-redshift \citep[$z=3.288$, ][]{2013AJ....145...69L}\footnote{Another, slightly different value for the redshift can also be found in the literature \citep[$z=3.20$,][]{1993ESOSR..13....1V}} quasar PMN~J0909+0354 \citep[hereafter J0909+0354; right ascension $9^\mathrm{h}9^\mathrm{min}15\fs91130$, declination $3\degr54\arcmin42\farcs7583$,][]{2021AJ....161...14P}
is a known radio and X-ray source. Here we present a study of its kpc- and pc-scale radio structure as well as \textit{Chandra} X-ray imaging of its kpc-scale emission.
In Section~\ref{target}, we introduce the target source. Section~\ref{observations} gives details of the radio and X-ray observations used in the analysis. Our results are presented in Section~\ref{results} and discussed in Section~\ref{discussion}. The paper is concluded with a summary in Section~\ref{summary}. 

For calculations, we applied parameters of the standard flat $\Lambda$CDM cosmological model as $H_0=70$~km~s$^{-1}$~Mpc$^{-1}$, $\Omega_\mathrm{M}=0.3$, and $\Omega_\Lambda=0.7$. At the redshift of the quasar, 1 milli-arcsecond (mas) angular separation corresponds to $7.481$~pc projected linear distance.

\section{The Target Quasar}
\label{target}

\subsection{Radio and X-ray Emission of J0909+0354}
\begin{table}[b]
    \centering
        \caption{Archival radio observations of J0909+0354.}    \label{tab:radioobs}
    \begin{tabular}{cccc}
    \hline\hline
              & $\nu$ (GHz)   & $S$ (mJy) & Reference \\
    \hline
    FIRST               & $1.4$  & $134.5\pm0.14$      & \citet{2015ApJ...801...26H}\\
    NVSS                & $1.4$  & $113.6\pm{3.4~~}$   & \citet{1998AJ....115.1693C} \\
    \hline
    \multirow{3}{*}{GBT}& $1.4$  & $213$                & \citet{1992ApJS...79..331W} \\
                        & $4.9$  & $123$                & \citet{1991ApJS...75....1B}\\
                        & $4.9$  & $111\pm11$           & \citet{1996ApJS..103..427G}\\
    \hline                    
    PMN                 & $5$    & $127\pm12$           & \citet{1995ApJS...97..347G}\\
    CLASS               & $8.4$  & $137.5$              & \citet{2003MNRAS.341....1M}\\
    \hline
    \end{tabular}
\end{table}

The quasar J0909+0354 has been detected with $\sim 100$-mJy level flux densities at various frequencies in different sky surveys. Flux densities from the following observations are listed in Table~\ref{tab:radioobs}.
Karl G. Jansky Very Large Array (VLA) A-configuration imaging observations at 1.5, 4.9, and 8~GHz revealed that the radio emission of the quasar can be resolved into a double structure: a compact core and a secondary component at about $2\arcsec$ angular separation in the north-northwestern direction \citep[][]{2014arXiv1406.4797G}. Archival Very Long Baseline Array (VLBA) observations \citep[][project code: BP171AB, PI: L. Petrov]{2021AJ....161...14P} of J0909+0354 at 4.3 and 7.6~GHz  show compact, unresolved radio emission up to $\sim10-20$~pc, with 111 and 76~mJy total flux densities, respectively\footnote{From \url{http://astrogeo.org/cgi-bin/imdb_get_source.csh?source=J0909\%2B0354}}. Global VLBI observations at 5~GHz resolved the pc-scale morphology, revealing a compact synchrotron self-absorbed core and a more diffuse jet structure visible up to $\sim10$~pc \citep{1999A&A...344...51P}.

\begin{figure}
\centering
\includegraphics[width=\linewidth]{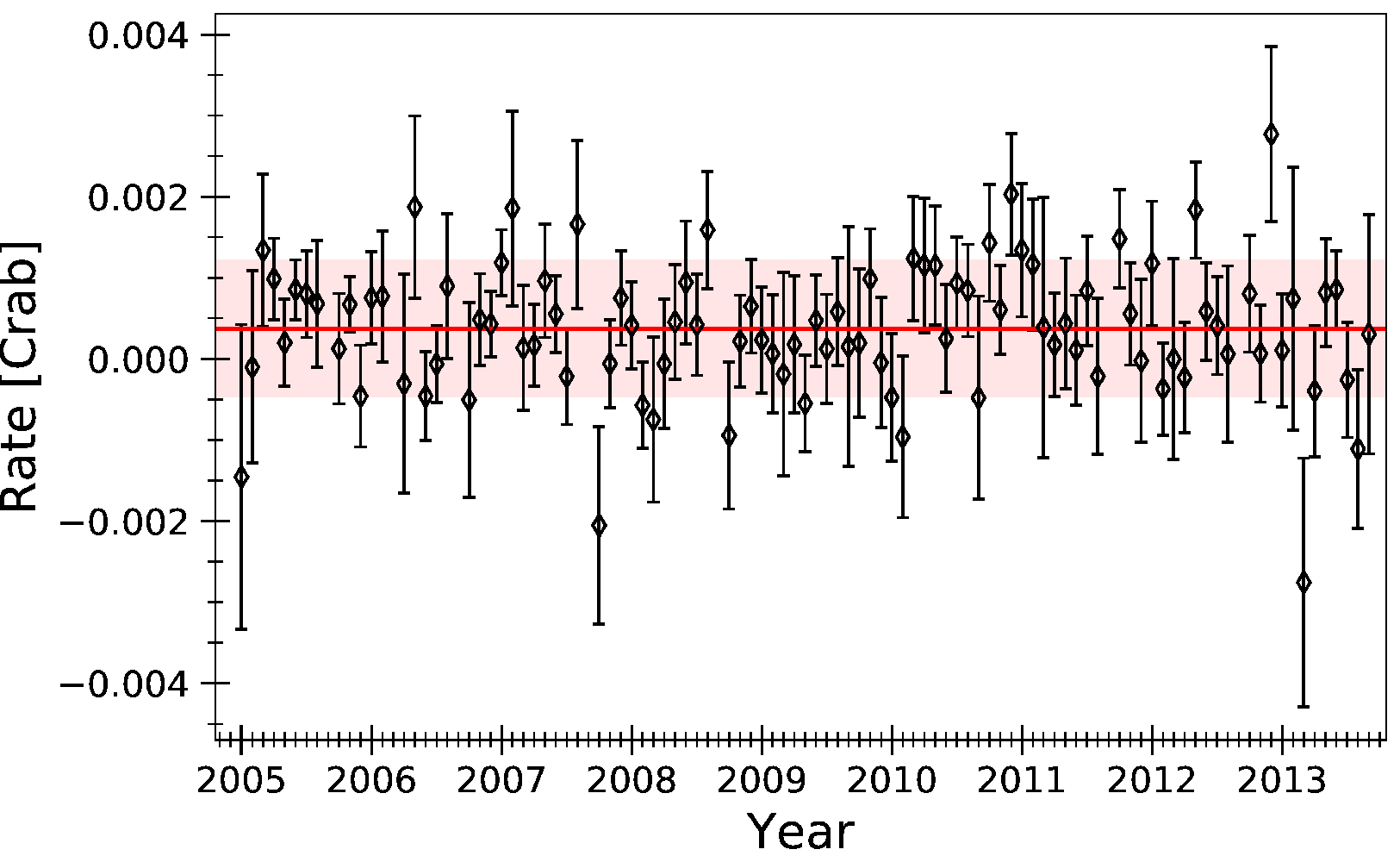}
\caption{Crab-weighted X-ray light curve of the quasar J0909+0354 based on the observations of the 105-month \textit{Swift}-BAT hard X-ray survey ($14-195$~keV). 
The total Crab flux in this energy range is $2.3343\times10^{-8}$~~~erg\,cm$^{-2}$\,s$^{-1}$, data were obtained from \url{http://swift.gsfc.nasa.gov/results/bs105mon/445}
\citep{2018ApJS..235....4O}. Unreliable data points (rates with relative uncertainties above $325\%$) were omitted from the plot. The red line and area denote the mean value and the standard deviation of the data points, respectively.}\label{fig:xlc}
\end{figure}

X-ray emission of the quasar was first detected by \textit{ROSAT} (ROentgen SATellite) in the $2-4$~keV energy range with a flux of $F=9.7\pm2.7\times10^{-13}$~~erg\,cm$^{-2}$\,s$^{-1}$ \citep{1997A&A...319..413B}. Fluxes from observations by the \textit{BeppoSAX} and \textit{Swift} space telescopes are $1.9\times10^{-12}$~~erg\,cm$^{-2}$\,s$^{-1}$ \citep[photon index $\gamma=1.16\pm0.2$,][]{2005A&A...433.1163D} and   $15.65\times10^{-12}$~~erg\,cm$^{-2}$\,s$^{-1}$ \citep[$\gamma=1.88_{-2.28}^{+1.51}$,][]{2018ApJS..235....4O}, in the $0.1-50$~keV and $14-195$~keV energy ranges, respectively. The light curve of J0909+0354 observed in the framework of the 105-month \textit{Swift}-BAT all-sky hard X-ray survey revealed some variability in the $14-195$~keV energy range (Fig.~\ref{fig:xlc}). 

\subsection{Emission in Other Wavebands}

\begin{table*}[!ht]
    \centering
        \caption{Archival ultraviolet-to-infrared photometry data of J0909+0354.}
    \label{tab:uvtoir}
    \begin{tabular}{ccccc}
    \hline\hline
          & Filter & $\lambda$& $m$~(mag) & Reference \\
\hline    
    \multirow{2}{*}{\textit{GALEX}}      & $FUV$   & 152.8~nm      &  $21.71$   & \multirow{2}{*}{\citet{2005ApJ...619L...1M,2020MNRAS.493.2745V}}\\
                                & $NUV$   & 227.1~nm      &  $21.83$   \\
 \hline                                
      \textit{Swift}-UVOT       & $u$     & 345.6~nm      &  $21.68$  & {\citet{2005SSRv..120...95R,2014ApSS.354...97Y}}  \\ 
 \hline
    \multirow{5}{*}{SDSS DR16}  & $u$     & 354.3~nm      &  $21.31\pm0.10$ &\multirow{5}{*}{\citet{2020ApJS..249....3A}} \\
                                & $g$     & 477.0~nm      &  $20.33\pm0.03$  \\
                                & $r$     & 623.1~nm      &  $19.91\pm0.02$  \\
                                & $i$     & 765.5~nm      &  $19.77\pm0.02$  \\
                                & $z$     & 913.4~nm      &  $19.80\pm0.08$  \\ 
 \hline    
    \multirow{3}{*}{\textit{Gaia} DR2}   &$G_\mathrm{BP}$&   500~nm      &  $20.52$  &\multirow{3}{*}{\citet{{2016AnA...595A...2G,2018AnA...616A...1G}}} \\
                                &$G$            &   700~nm      &  $20.02$  \\
                                &$G_\mathrm{RP}$&   850~nm      &  $19.51$  \\    
 \hline  
    \multirow{5}{*}{Pan-STARRS} & $g$     & 486.6~nm      &  $20.24\pm0.02$&\multirow{5}{*}{\citet{2016arXiv161205560C}}  \\
                                & $r$     & 621.5~nm      &  $19.94\pm0.02$  \\
                                & $i$     & 754.5~nm      &  $19.82\pm0.02$  \\
                                & $z$     & 867.9~nm      &  $19.77\pm0.04$  \\
                                & $y$     & 963.3~nm      &  $19.58\pm0.04$  \\
 \hline
    \multirow{4}{*}{UKIDSS}     & $Y$     & 1.031~$\mu$m  &  $19.19\pm0.06$ &\multirow{4}{*}{\citet{2007MNRAS.379.1599L}} \\
                                & $I$     & 1.248~$\mu$m  &  $18.64\pm0.07$  \\
                                & $H$     & 1.630~$\mu$m  &  $18.26\pm0.14$  \\
                                & $K$     & 2.201~$\mu$m  &  $16.99\pm0.07$  \\
 \hline
    \multirow{4}{*}{\textit{WISE}}    & $W1$    & ~3.35~$\mu$m  &  $16.30\pm0.07$ &\multirow{4}{*}{\citet{2010AJ....140.1868W,2014yCat.2328....0C}} \\
                                & $W2$    & ~4.60~$\mu$m  &  $15.62\pm0.14$  \\
                                & $W3$    &  11.6~$\mu$m  &  $11.63\pm0.25$  \\
                                & $W4$    &  22.1~$\mu$m  &  ${~~8.81}\pm0.52$  \\  
    \hline  
    \end{tabular}

\end{table*}

The quasar J0909+0354 is also a source of electromagnetic radiation detected in various surveys in the ultraviolet, optical, and infrared wavebands, which are listed in  Table~\ref{tab:uvtoir}.

The quasar was not detected in $\gamma$-rays with either of the \textit{INTErnational Gamma-Ray Astrophysics Laboratory} \citep[\textit{\textit{INTEGRAL}}, ][]{2003A&A...411L...1W}, the Large Area Telescope of  \textit{Fermi Gamma-Ray Space Telescope} \citep[\textit{Fermi}-LAT][]{2009ApJ...697.1071A}, the
\textit{Compton Gamma Ray Observatory} \citep[\textit{CGRO}, e.g.][]{1988SSRv...49...69K,1998AIPC..428....3M,1993ApJS...86..629T,2013ApJS..208...21G}, or the  \textit{Astro-rivelatore  Gamma  aImmagini LEggero} mission \citep[\textit{AGILE}, ][]{2009A&A...502..995T}.

\section{Observations and Data Reduction}
\label{observations}

\subsection{Very Large Array}
For a quantitative comparison of radio emission between pc and kpc scales, we utilized data obtained with the VLA at 1.5, 6.2, and 8.5~GHz.

The 8.5~GHz observations were carried out on 1998 March 14, in the framework  of the Cosmic Lens All-Sky Survey \citep[CLASS,][]{2003MNRAS.341....1M} (project code: AM593, PI: S. Myers), in which 27 stations participated. The on-source time for J0909+0354 was 39~s. Two intermediate frequency channels (IFs) were used with one spectral channel in each, the total bandwidth was 50~MHz. The data were recorded in full polarization with 3.3~s integration time. We calibrated the phases and amplitudes with the National Radio Astronomy Observatory (NRAO) Astronomical Image Processing System\footnote{\url{http://www.aips.nrao.edu/index.shtml}} \citep[\textsc{AIPS}, e.g.,][]{1995ASPC...82..227D,2003ASSL..285..109G} package, following the steps of standard data reduction described in the cookbook\footnote{\url{http://www.aips.nrao.edu/cook.html}} for VLA continuum data, using 3C~48 as primary flux density calibrator.

The $6.2$~GHz observations were conducted on 2012 November 18  (project code: 12B-230, PI: J. Wardle) with the participation of 26 telescopes. In the $3.47$-h long observing run, 39 sources were targeted including calibrators, from which the on-source time for J0909+0354 was 316~s. A total of 16 IFs were used with 64 spectral channels in each IF. The total bandwidth was $128$~MHz. The data were recorded in full polarization, and were correlated with 1~s averaging time.

The measurements at $1.5$~GHz were conducted on 2016 October 25 (project code: 16B-015, PI: J. S. Farnes), with 26 antennae participating. From the total 10~h of observation, the on-source time for J0909+0354 was 126~s. The 64~MHz total bandwidth was divided into 16 IFs, with 64 spectral channels in each. The data were recorded in full polarization, and were correlated with 1~s averaging time.

We calibrated the phases and amplitudes of the 1.5 and 6.2~GHz VLA data sets  with the Common Astronomy Software Applications\footnote{\url{http://casa.nrao.edu}} \citep[\textsc{casa}, ][]{2007ASPC..376..127M} software using 3C~48, J0738+1741, and J0831+0429 as calibrators, following the steps of the standard data reduction for VLA continuum observations\footnote{\url{http://casaguides.nrao.edu/index.php?title=VLA_Continuum_Tutorial_3C391-CASA5.5.0}}. Then we exported the data to \textsf{uvfits} format. 

The calibrated visibilities of all three VLA observations were then imported to the \textsc{difmap} program\footnote{\url{ftp://ftp.astro.caltech.edu/pub/difmap/difmap.html}} \citep{1997ASPC..125...77S}, where we carried out hybrid mapping with cycles of phase and amplitude self-calibration and imaging \citep[applying the \textsf{clean} deconvolution method by][]{1974A&AS...15..417H}. Finally, to quantitatively describe the brightness distribution of the source, we fitted circular Gaussian model components directly to the self-calibrated visibility data \citep{1995ASPC...82..267P}. Uncertainties for the fitted model parameters were calculated following \citet{2008AJ....136..159L}.

\subsection{Very Long Baseline Interferometry}\label{sec:vlbidata}
To study the pc-scale radio structure of the quasar J0909+0354, we used archival data as well as new observations made by various VLBI arrays. The latest and most sensitive data set was acquired at 5~GHz with the European VLBI Network (EVN) on 2019 March 1 (project code: EP115, PI: K. Perger). The observation lasted for a total 6~h involving 15 radio telescopes: Jodrell Bank Mk2 (United Kingdom), Westerbork (The Netherlands), Effelsberg (Germany), Medicina, Noto (Italy), Onsala (Sweden), Tianma, Nanshan (China), Toru\'n (Poland), Yebes (Spain),
Svetloe, Zelenchukskaya, Badary (Russia), Hartebeesthoek (South Africa), and Irbene (Latvia). The on-source integration time was 5.24~h. The data were recorded at a rate of $1024$~Mbit~s$^{-1}$ in left and right circular polarizations. The total bandwidth was 16~MHz per polarization in 32 spectral channels per IF, and a total of 8 IFs were used. The data were correlated with 2~s averaging time at the EVN Data Processor at the Joint Institute for VLBI European Research Infrastructure Consortium (Dwingeloo, the Netherlands). 

We calibrated the phases and amplitudes of the visibilites in the \textsc{AIPS} package. We applied a priori amplitude calibration based on radio telescope gain curves and measured system temperatures, then removed inter-channel delay and phase offsets  using a 1-min data segment of a bright calibrator source (J0909+0121). Visual inspection and flagging of the visibilities were followed by global fringe fitting \citep{1983AJ.....88..688S} on the target source J0909+0354. The calibrated data were exported into \textsf{uvfits} format for further analysis.

For a comparison, we also recovered and analyzed the archival data observed by a global VLBI network on 1992 September 27--28 \citep{1999A&A...344...51P}. Nine telescopes, Effelsberg, Medicina, Onsala, the phased array of the Westerbork Synthesis Radio Telescope, as well as Green Bank, Haystack, Owens Valley, and the phased VLA (the latter four in the USA) participated in the observations which were part of a 24-h long  experiment. The on-source time for J0909+0354 was 3~h. The data were recorded in left circular polarization with a total bandwidth of 28~MHz in 7~IFs, and were correlated in the Max Planck Institute for Radio Astronomy (Bonn, Germany). For further analysis, we used the visibility data calibrated by \citet{1999A&A...344...51P}.

To supplement our long-track VLBI observations for studying possible changes in the pc-scale radio structure of J0909+0354, we also analyzed archival `snapshot' data obtained with the VLBA. These observations were conducted in the framework of the 7th VLBA Calibrator Survey\footnote{\url{http://astrogeo.org/vcs7/}} \citep[][project code: BP171AB, PI: L. Petrov]{2021AJ....161...14P}  on 2013 April 28. All ten telescopes of the array participated in the dual-frequency (4.3 and 7.6~GHz) observation which was carried out in right circular polarization, with an on-source time of 1~min. The total bandwidth was 32~MHz in 8~IFs. The a priori calibrated visibility data sets were produced by the \textsc{pima} v2.03 software \citep{2011AJ....142...35P} and were obtained from the Astrogeo VLBI image database\footnote{\url{http://astrogeo.org/}}.

All four calibrated  VLBI data sets were then imported to the \textsc{difmap} program for phase and amplitude self-calibration, imaging and model fitting, similarly to the VLA data treatment described above. Errors for the fitted model parameters were calculated following \citet{2008AJ....136..159L}, considering an additional $5\%$ calibration uncertainty for flux densities.

\subsection{X-ray Observations}
The X-ray emission associated with the quasar J0909+0354 was observed with the \textit{Chandra} Advanced CCD Imaging Spectrometer (ACIS) as part of a survey of radio-loud quasars at $z>3$ (ObsID 20404, PI: D. Schwartz). We then used 77.5-ks follow-up observation (ObsIDs 22568, 23161, and 23162, PI: D. Schwartz) to reveal the extended X-ray features. The latter observations took place on February 17, 18, and 20, in 2020. The data were reduced with \textsc{ciao} (\textit{Chandra} Interactive Analysis of Observations) version 4.12 \citep{2006SPIE.6270E..1VF}, using \textsc{Sherpa} version 2 \citep{2007ASPC..376..543D}. Imaging used \textsc{SAOImage ds9} version 8.2b1 \citep{2005ASPC..347..110J}. Background was determined to be 0.0591$\pm$0.0013~counts\,arcsec$^{-2}$ from two rectangular regions totaling 34,131 arcsec$^2$, away from the quasar and not including sources.

\section{Results}
\label{results}

\subsection{Kiloparsec-scale Structure}
\label{ss:kpc-str}

\begin{figure*}[!h]
    \centering
    \gridline{
    \fig{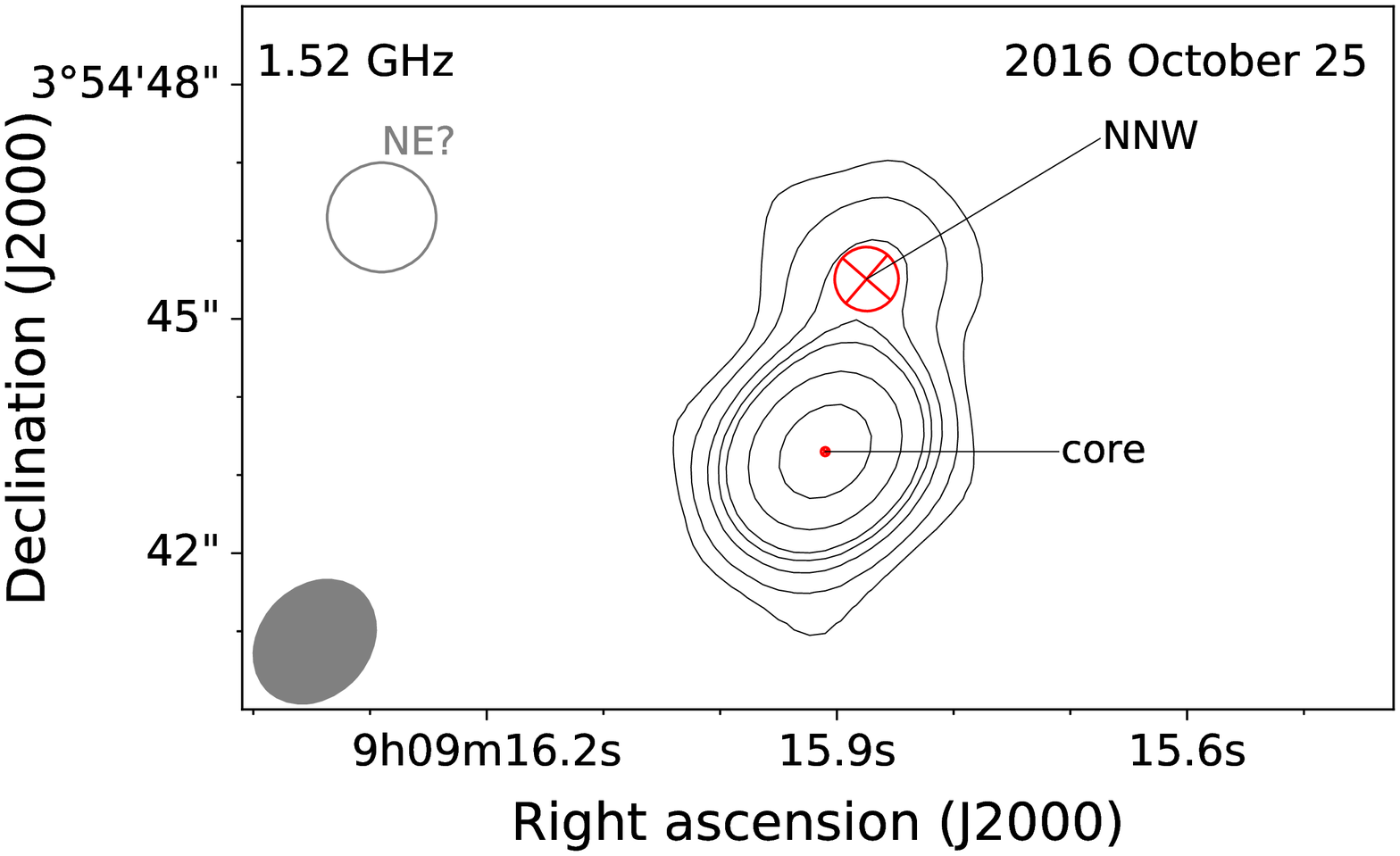}{0.5\linewidth}{(a) Naturally weighted clean map at 1.5~GHz from the VLA observations. The maximum intensity is 142~mJy~beam$^{-1}$. The rms noise is 0.75~mJy~beam$^{-1}$. The restoring beam is $1\farcs4\times1\farcs7$ (at $-43\degr$ position angle).}}
     \gridline{
    \fig{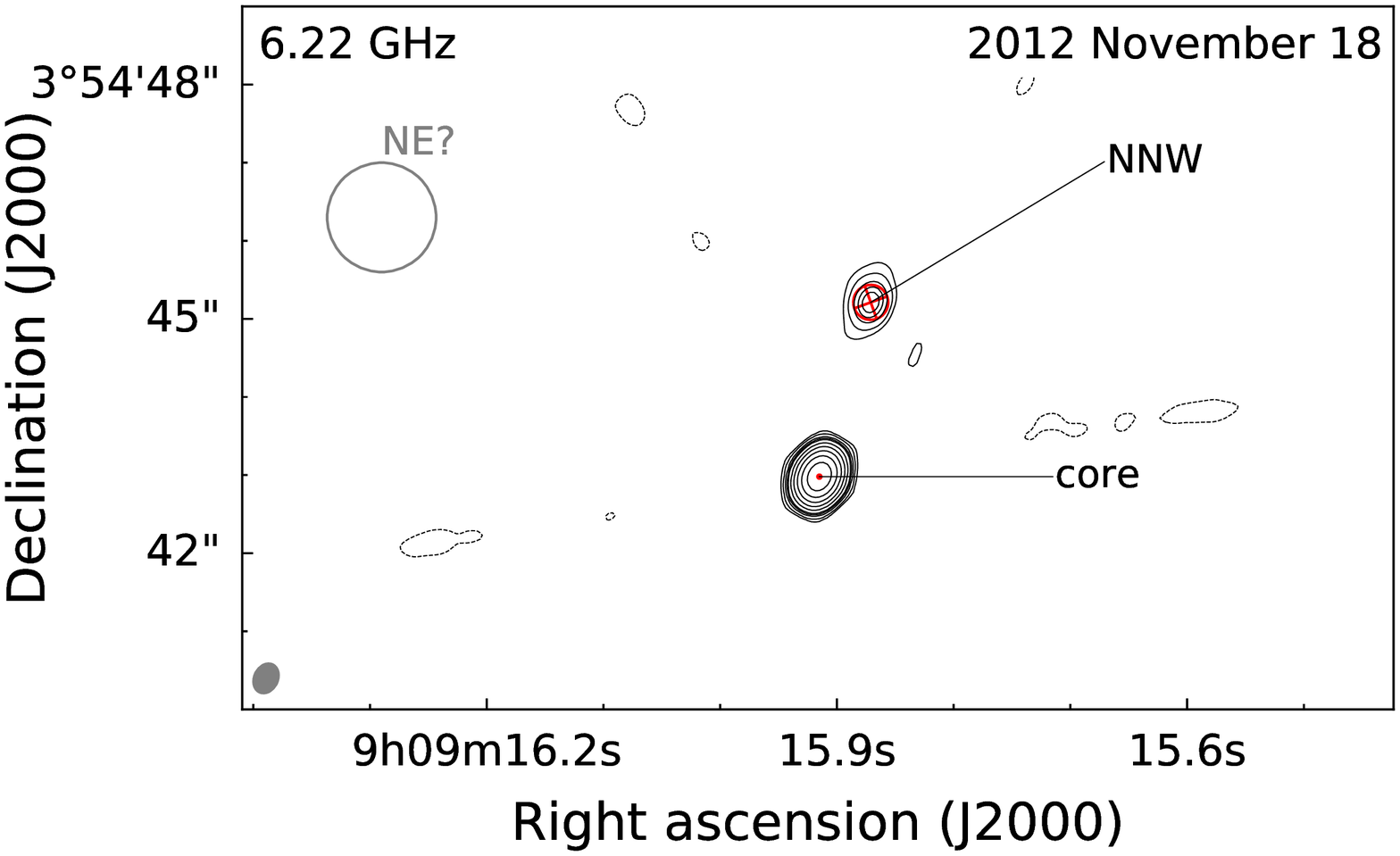}{0.5\linewidth}{(b) Naturally weighted clean map at 6.2~GHz from the VLA observations. The maximum intensity is 190~mJy~beam$^{-1}$. The rms noise is 0.13~mJy~beam$^{-1}$. The restoring beam is $0\farcs3\times0\farcs4$ (at $-26\degr$ position angle).}}
     \gridline{
    \fig{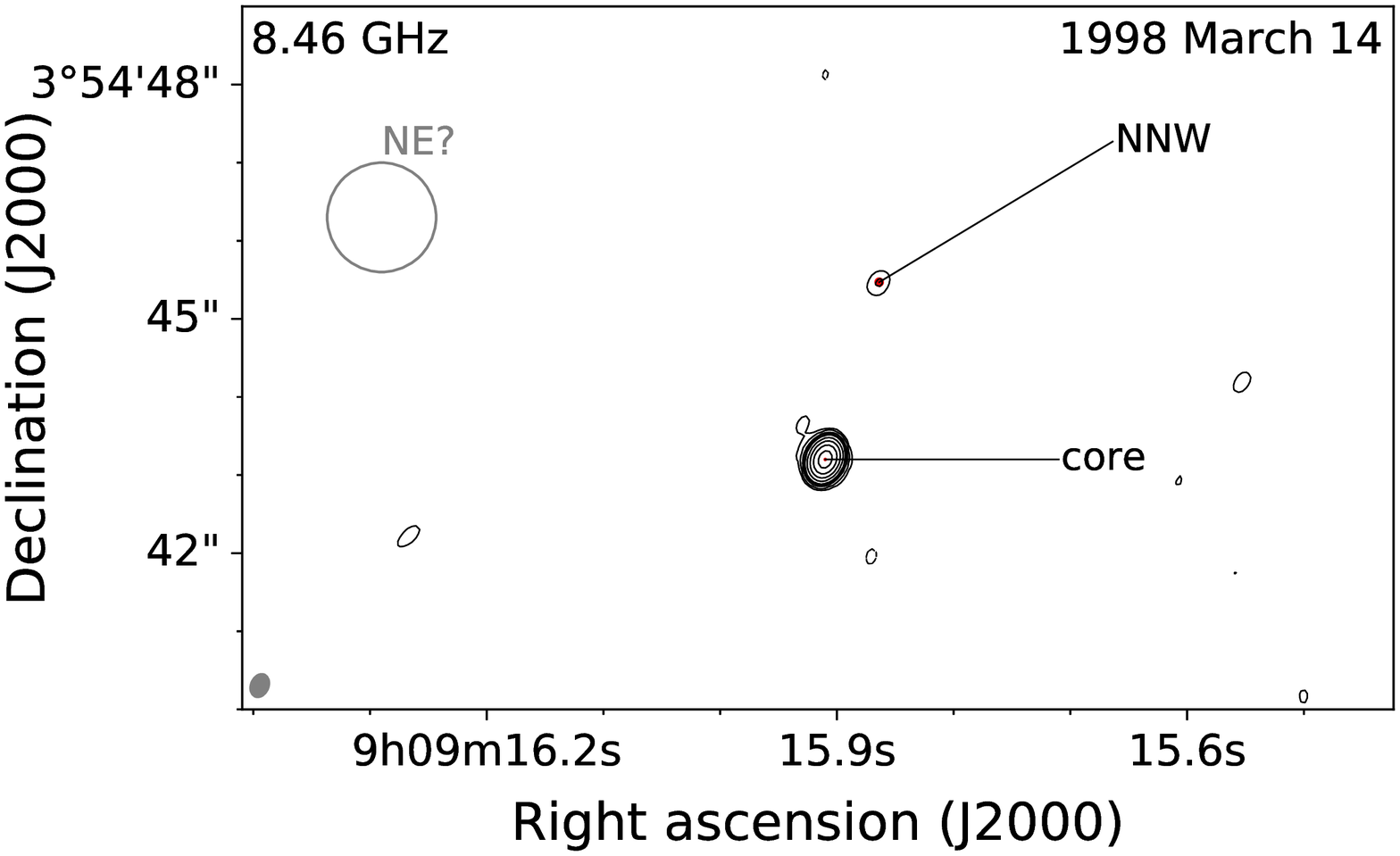}{0.5\linewidth}{(c) Naturally weighted clean map at 8.5~GHz from the VLA obsrvations. The maximum intensity is 208~mJy~beam$^{-1}$. The rms noise is  0.32~mJy~beam$^{-1}$. The restoring bean is $0\farcs2\times0\farcs3$ (at $-23\degr$ position angle).} }
     \caption{Naturally weighted clean maps of J0909+0354 from the VLA observations. The first contours are drawn at $\pm3$ times the rms noise, the positive levels increase by a factor of 2. The fitted model components are denoted with red crossed circles. The restoring beams are shown in the bottom left corners.  The light gray circle in the upper left part of the maps marks the position of the NE component identified in the \textit{Chandra} image (Fig.~\ref{fig:J0909xray}) showing no radio emission here. }\label{fig:vla}
\end{figure*}

Model fitting to the visibilities of the 1.5, 6.2, and 8.5-GHz VLA observations resulted in two components for all three data sets: a compact core and a slightly more diffuse feature (discussed in detail in subsection \ref{ss:nnw}) at $\sim2\farcs33$ to the north-northwestern (NNW) direction (position angle  $\sim-16\fdg5$ as measured from north through east, Fig.~\ref{fig:vla}). The sums of the flux densities of the circular Gaussian model components describing the kpc-scale radio structure is $153.1\pm14.4$~mJy, $198.9\pm12.6$~mJy, and $211.1\pm13.3$~mJy, for the 1.5, 6.2, and 8.5~GHz data, respectively\footnote{Although we utilized the same 8.5-GHz data set, our data reduction resulted in a flux density $30\%$ higher than reported by \citet[][Table~\ref{tab:radioobs}]{2003MNRAS.341....1M}. As we found the same scaling factor for all other sources in the observation, we attribute this to the different amplitude calibration process in \textsc{AIPS}.}. The properties of the two fitted components for the VLA data sets are listed in Table~\ref{tab:J0909_vlamodel}. As it was previously noted by \citet{2014arXiv1406.4797G}, we examined the nature of NNW, and found that its three-point spectral index\footnote{Used in the $S_\nu\propto\nu^\alpha$  convention.} between 1.5 and 8.5~GHz is $\alpha_\mathrm{NNW}=-1.08\pm 0.17 $. The three-point spectral index for the core between 1.5 and 8.5~GHz is $\alpha_\mathrm{core}=0.19\pm0.01$ To avoid the possible effect of time variability, we also determined the spectral index for this component by separately processing the first and last two IFs from the same 6.2~GHz observation. Repeating the hybrid mapping and Gaussian model fitting on these two subsets, we found the values $\alpha_\mathrm{7.77,core}^\mathrm{4.62}=-0.51\pm 0.05$ and  $\alpha_\mathrm{7.77,NNW}^\mathrm{4.62}=-1.03\pm 0.10$ for the core and NNW components, respectively. The latter value is consistent with the identification of the NNW component as a steep-spectrum jet hotspot -- see further discussion on this in subsection \ref{ss:nnw} below. 

\begin{table*}[t]
\centering
\caption{Model parameters of J0909+0365 from the VLA observations}\label{tab:J0909_vlamodel}
\begin{tabular}{cccccccc}
\hline\hline
 	&& \multirow{2}{*}{R.A.} & \multirow{2}{*}{Dec.}  & \multirow{2}{*}{$\vartheta$} & $S$ & $R$ & $\phi$\\
&&&&&(mJy)& ($''$)& ($\degr$)\\
\hline
\multirow{2}{*}{1.5~GHz}& core & $9^\mathrm{h}9^\mathrm{min}15\fs91\pm0\fs001$ & $3\degr54'43\farcs3\pm0\farcs01$ & $<0\farcs013\pm0\farcs01$ & $133.1\pm13.75$ &0 &0\\
&NNW			& $9^\mathrm{h}9^\mathrm{min}15\fs87\pm0\fs02\phn$ & $3\degr54'45\farcs5\pm0\farcs11$ & $0\farcs82\pm0\farcs22$& $20.0\pm4.4$ &$2.27\pm0.11$& $-13.5\pm2.7$\\
\hline
\multirow{2}{*}{6.2~GHz}&core 		& $9^\mathrm{h}9^\mathrm{min}15\fs91\pm0\fs001$ & $3\degr54'43\farcs0\pm0\farcs01$ & $0\farcs048\pm0\farcs002$ & $193.6\pm12.5$ &0 &0\\
&NNW			& $9^\mathrm{h}9^\mathrm{min}15\fs87\pm0\fs04\phn$ & $3\degr54'45\farcs2\pm0\farcs07$ & $0\farcs45\pm0\farcs13$& $5.3\pm1.2$ &$2.33\pm0.04$& $-16.5\pm1.0$\\
\hline
\multirow{2}{*}{8.5~GHz}&core &	$9^\mathrm{h}9^\mathrm{min}15\fs91\pm0\fs001$&$3\degr54'43\farcs2\pm0\farcs01$ & $<0\farcs010\pm0\farcs01$ & $208.3\pm13.25$ &0 &0\\
&NNW			& $9^\mathrm{h}9^\mathrm{min}15\fs86\pm0\fs01\phn$ & $3\degr54'45\farcs5\pm0\farcs02$ & $<0\farcs09\pm0\farcs04$&$ 2.8\pm1.0$ &$2.37\pm0.02$& $-17.0\pm0.4$\\
\hline
\end{tabular}
 \tablecomments{Column 1 -- model component name, Columns 2, 3 -- coordinates (right ascension, R.A., and declination, Dec.), Column 4 -- circular Gaussian model component size (FWHM); due to its small size, the uncertainty of the core component is given as the relative astrometric precision of the VLA \citep[calculated following the error estimation of][]{2018arXiv180505266B}. We note that these uncertainties are in the same order of magnitude as the values calculated following the VLA observation guide (\url{http://science.nrao.edu/facilities/vla/docs/manuals/oss/performance/positional-accuracy}), i.e. $10\%$ of the FWHM of the restoring beam, Column 5 -- flux density, Column 6 -- radial distance from the core component, Column 7 -- model position angle with respect to the core measured from north through east.}
\end{table*}

\begin{figure}
\centering
\includegraphics[width=\linewidth]{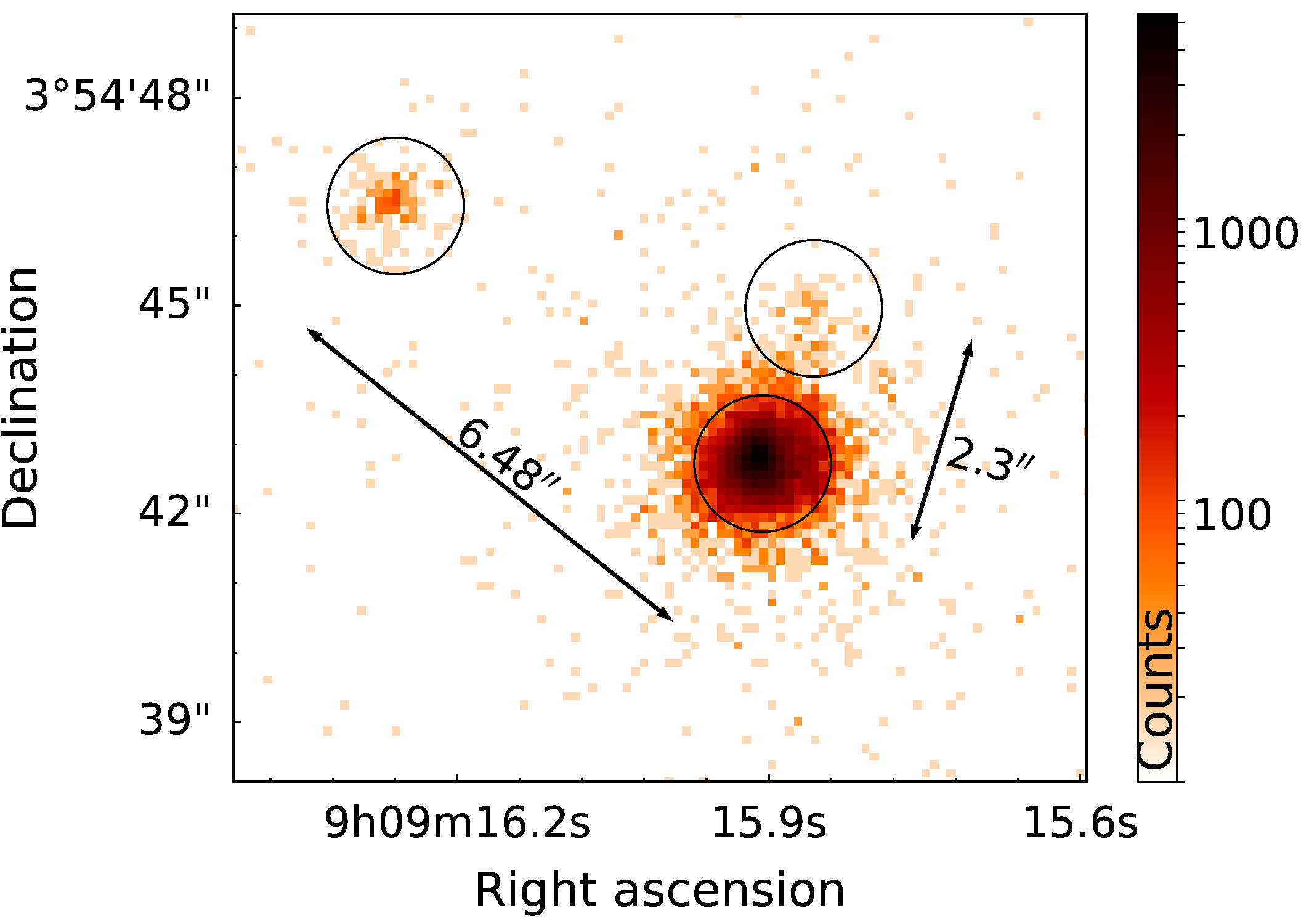}
\caption{\textit{Chandra} X-ray image of J0909+0354. Photons are binned in 1/4 ACIS pixels ($0\farcs123$) in logarithmic intervals from 1 to 335 counts (see the color scale for counts per ACIS pixel). Circles of $1\arcsec$ radius are drawn about the quasar, the NE source, and the NNW feature.}\label{fig:J0909xray}
\end{figure}

The \textit{Chandra} 0.5$-$7 keV X-ray image of J0909+0354 is shown in Fig.~\ref{fig:J0909xray}, and reveals three distinct features. There is bright emission from the quasar, from a point-like source in $6\farcs48$ angular distance at position angle $55\degr$, and emission extending from the quasar $2\farcs3$ toward  the  NNW component at position angle $-17\fdg4$. We interpret the latter feature as a kpc-scale jet, with enhanced emission at its end coincident with NNW. 

To establish the reality of the jet, we performed a high fidelity simulation of the quasar, using saotrace-2.0.4\_03\footnote{\url{http://cxc.harvard.edu/cal/Hrma/SAOTrace.html}} to generate rays which are passed to
marx-5.5.0\footnote{\url{https://space.mit.edu/CXC/MARX/}} \citep{2012SPIE.8443E..1AD} to simulate an
ACIS-S image. We measure 9539 counts in a 1\farcs4 circle around the quasar, 54 in a 2\farcs4 $\times$0\farcs6 box region defined between the core and NNW, and 28 in the circle around the NNW radio emission at the end of the jet. Normalizing the simulated counts to the number in the quasar (i.e. the core component), we predict that 36.5 counts from the quasar scatter into the jet box, and that 6.4 scatter into the NNW region, giving a net 17.2$\pm$9.5 counts for the box area, and 21.5$\pm$5.9 in the  NNW region. Interpreting all the emission to the NNW as a single jet would result in 39.0$\pm$10.4 counts. With the average background of 0.03 counts, detections of X-ray photons are significant ($\geq3\sigma$) in the box and NNW regions, either considering the emission as an ensemble continuous jet or as two separate features. Taking the dimensions of the box area and the $0\farcs4$ extraction circle around NNW into account, there are $11.9$~counts~arcsec$^{-2}$ and $42.8$~counts~arcsec$^{-2}$ in the two features, respectively, implying an enhancement at the latter component.

The quasar fits a power law photon spectrum $\propto E^{-1.24\pm0.03}$, using the galactic absorption $n_H = 3.5 \times 10^{20}$~cm$^{-2}$ and also fitting an intrinsic absorption of $(2.7^{+1.8}_{-1.7}) \times 10^{22}$~cm$^{-2}$ at the source. The measured $0.5-7$~keV flux of $1.76 \times 10^{-12}$~erg\,cm$^{-2}$\,s$^{-1}$ corresponds to a luminosity of $1.78 \times 10^{47}$~erg\,s$^{-1}$ in the 2.1 to 30~keV rest frame band. This value is consistent with previous X-ray measurements in similar bandwidths. Interpreting all the emission to the NNW as due to an X-ray jet, it would have a photon spectrum $\propto E^{-1.71\pm0.30}$ and a flux of $(8.5\pm2.5) \times 10^{-15}$~erg\,cm$^{-2}$\,s$^{-1}$. Such a jet flux would be $0.5\%$ of the quasar flux. This is consistent with but somewhat lower than the median $2\%$ jet-to-core ratio from a sample of quasars at lower redshifts \citep{2018ApJ...856...66M}. We cannot exclude a significant contribution to the jet flux from a point-like X-ray source at the NNW component. Taking the extraction circle of $0\farcs4$ radius about the NNW region, we fit a photon spectrum  $\propto E^{-1.42\pm0.44}$. Considered as an upper limit, such a source could be contributing $4.5 \times 10 ^{-15}$~erg\,cm$^{-2}$\,s$^{-1}$ of the flux extending to the NNW. 

The source to the NE fits a power law with photon index $1.81 \pm 0.33$, giving a flux $2.03 \times 10 ^{-14}$~erg\,cm$^{-2}$\,s$^{-1}$.

\subsection{Parsec-scale Structure}
\begin{figure*}
\gridline{
\fig{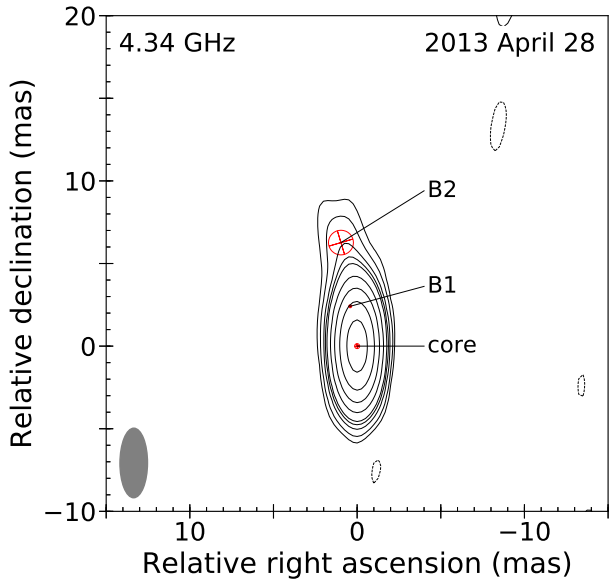}{0.45\textwidth}{(a) Naturally weighted clean map at 4.3~GHz from the VLBA observations.  The maximum intensity is $101.7$~mJy~beam$^{-1}$, the rms noise is $0.2$~mJy~beam$^{-1}$, the restoring beam is $1.7$~mas$\times4.2$~mas (at $0\degr$ position angle).}
\fig{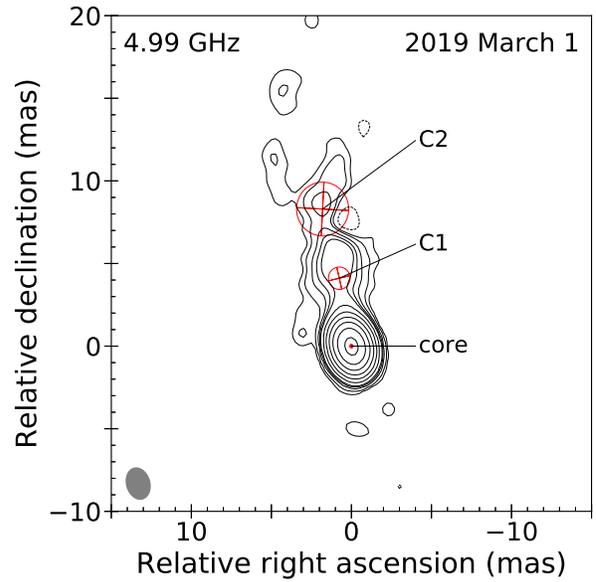}{0.45\textwidth}{(b) Naturally weighted clean map at 5~GHz from the 2019 EVN observations.
The maximum intensity is $64.09$~mJy~beam$^{-1}$, the rms noise is $0.03$~mJy~beam$^{-1}$, the restoring beam is $1.4$~mas$\times1.9$~mas (at $18\degr$ position angle).}
}
\gridline{
\fig{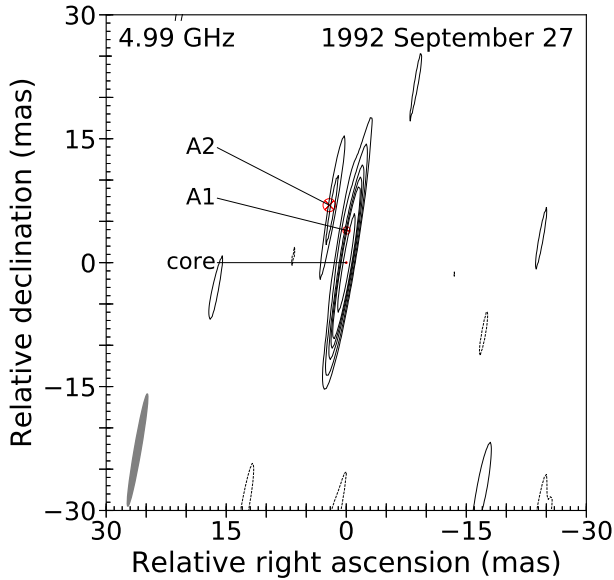}{0.45\textwidth}{(c) Naturally weighted clean map at 5~GHz from the global VLBI observations \citep{1999A&A...344...51P}. The maximum intensity is $36.98$~mJy~beam$^{-1}$, the rms noise is $0.47$~mJy~beam$^{-1}$, the restoring beam is $0.8$~mas$\times13.8$~mas (at $-10\degr$ position angle).}
\fig{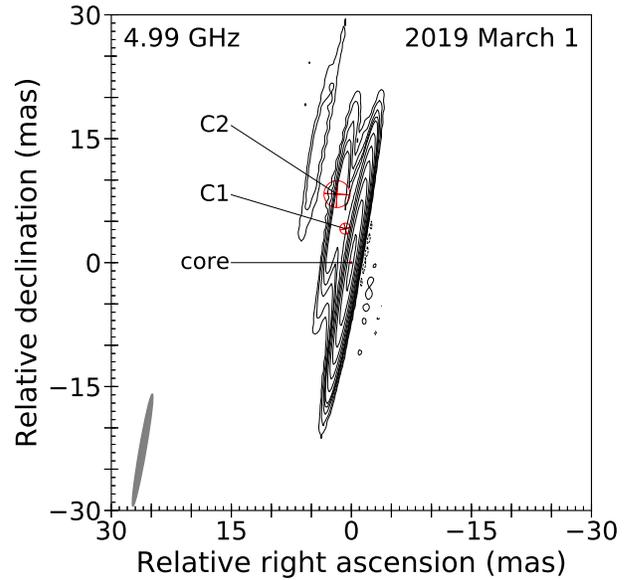}{0.45\textwidth}{(d) Naturally weighted clean map at 5~GHz from the EVN observations restored with the restoring beam of the global VLBI observations in 1992  ($0.8$~mas$\times13.8$~mas, at $-10\degr$ position angle). The maximum intensity is $64.09$~mJy~beam$^{-1}$, the rms noise is $0.03$~mJy~beam$^{-1}$.}
}
\caption{Naturally weighted clean maps of J0909+0354  at frequencies around 4 to 5~GHz. The first contours are drawn at $\pm3$ times the rms noise, the positive levels increase by a factor of 2. The fitted model components are denoted with red crossed circles. The Gaussian restoring beam is shown in the bottom left corners.\label{fig:J0909_vlbi}}
\end{figure*}

In the original publication of the global VLBI observations, \citet{1999A&A...344...51P} identified multiple jet components in the pc-scale radio structure of the quasar J0909+0354. However,  a conclusive analysis of the jet structure extended to the north was hampered by the lack of the long north--south baselines in the global VLBI array. This resulted in a relatively poorer angular resolution in that position angle. After our re-analysis of the 1992 data, the best Gaussian fit to the visibility data provided 3 model components (Fig.~\ref{fig:J0909_vlbi}c, Table~\ref{tab:J0909_model}). The integral flux density  based on the fitted circular Gaussian model components is $52.5\pm5.2$~mJy.  Since the major axis of the elongated restoring beam coincided with the jet direction, additional observations were needed for the reliable characterization of the inner jet structure in J0909+0354.

\begin{table*} 
\centering
\caption{Model parameters of J0909+0354 from the VLBI observations}\label{tab:J0909_model}
\begin{tabular}{ccccccccc}
\hline\hline
& & $S$ 		& $R$ 	&$\phi$	& $\vartheta$ 	&$T_\mathrm{B}$ & \multirow{2}{*}{$\delta_\mathrm{eq}$}& \multirow{2}{*}{$\delta_\mathrm{char}$}\\
&				 &(mJy) &(mas)	& ($\degr$) & (mas) 	&($10^{11}$~K) & \\
\hline
 global& core &$35.9\pm4.6$ & 0 & 0 &$<0.77\pm0.09$  &$>0.13\pm0.04$ &$>0.3$ &$>0.4$\\
1992& A1     &$11.9\pm2.2$  &$3.9\pm0.2$  &$-0.5\pm2.6$ &$<1.6\pm0.4$\\
  5 GHz                       & A2   &$4.7\pm1.3$  &$7.3\pm0.7$  &$16.9\pm5.2$ &$<3.0\pm1.3$\\
\hline
\multirow{3}{*}{4.3 GHz}    &core & $102.2\pm6.7$& 0 & 0 & $0.30\pm0.01$ & $ 3.2\pm0.5$ & $6$& $10$\\
                     &B1  & $3.9\pm2.5$ & $2.5\pm0.1$& $ 9.8\pm0.2$ & $<0.2\pm0.1$\\
 \multirow{2}{*}{VLBA}     &B2  & $2.8\pm0.7$ & $6.3\pm0.2$&$8.8\pm2.1$& $1.5\pm0.5$\\
\multirow{2}{*}{2013}      & \cline{1-8} 
\multirow{2}{*}{7.6 GHz}    &core & $67.4\pm4.8$& 0 & 0 & $0.17\pm0.01$ & $ 2.1\pm0.4$ & $4$& $7$\\
         		            &B  & $7.7\pm2.5$ & $1.25\pm0.01$& $6.4\pm0.6$& $0.35\pm0.02$\\
\hline
EVN    & core &$65.3\pm3.6$ & 0  & 0  &$0.22\pm0.01$  &$2.9\pm0.3$   &$6$ &$10$\\
2019  & C1   &$3.4\pm0.3$  &$4.2\pm0.1$  &$10.5\pm1.8$  &$1.4\pm0.3$\\
5 GHz        	            & C2   &$1.6\pm0.2$  &$8.5\pm0.4$  &$12.2\pm2.9$  &$3.3\pm0.9$\\
   \hline
\end{tabular}
\tablecomments{ Column 1 -- VLBI array, observing year, and frequency, Column 2 -- model component name, Column 3 -- flux density, Column 4 -- radial distance from the core component, Column 5 --  component position angle with respect to the core  measured from north through east, Column 6 -- circular Gaussian model component size (FWHM) or upper limit corresponding to the minimum resolvable angular size \citep{2005AJ....130.2473K}, Column 7 -- brightness temperature of the core component, Column 8 -- equipartition Doppler factor  \citep{1994ApJ...426...51R}, Column 9 -- Doppler factor calculated assuming $T_\mathrm{b,int}=3\times10^{10}$~K \citep{2006ApJ...642L.115H}. }
\end{table*}

The longer north--south interferometric baselines of the 2019 EVN observation resulted in a nearly circular restoring beam (Fig.~\ref{fig:J0909_vlbi}b). Three circular Gaussian components were found as the best-fit model of the brightness distribution (Table~\ref{tab:J0909_model}). The higher north--south resolution of the EVN observations allows  us to conduct a detailed analysis of the pc-scale morphology of the quasar. The jet propagates towards the north, and is clearly detectable up to $\sim20$~mas ($\sim 150$~pc). It becomes more diffuse with the increasing distance with respect to the core, apparently splitting into two branches like a fork at $\sim10$~mas. The peak intensity in the image is $64.1\pm1.5$~mJy~beam$^{-1}$, the integral flux density of the fitted model components is $70.2\pm3.6$~mJy (within $\sim65$~pc projected distance; Fig.~\ref{fig:J0909_vlbi}b, Table~\ref{tab:J0909_model}). 

For a qualitative comparison of the 1992 and 2019 images obtained at the same observing frequency, we restored the \textsf{clean} map of the most recent 2019 observations with the same elongated beam as of the 1992 global VLBI observations (Fig.~\ref{fig:J0909_vlbi}d). The overall structure of the source is clearly similar at both epochs, with more diffuse emission detected in the outer regions of the jet in 2019 (Fig.~\ref{fig:J0909_vlbi}c and \ref{fig:J0909_vlbi}d). This can be attributed to the higher sensitivity of the new EVN observations compared to the old global VLBI observations.

\begin{figure}
    \centering
    \includegraphics[width= \linewidth]{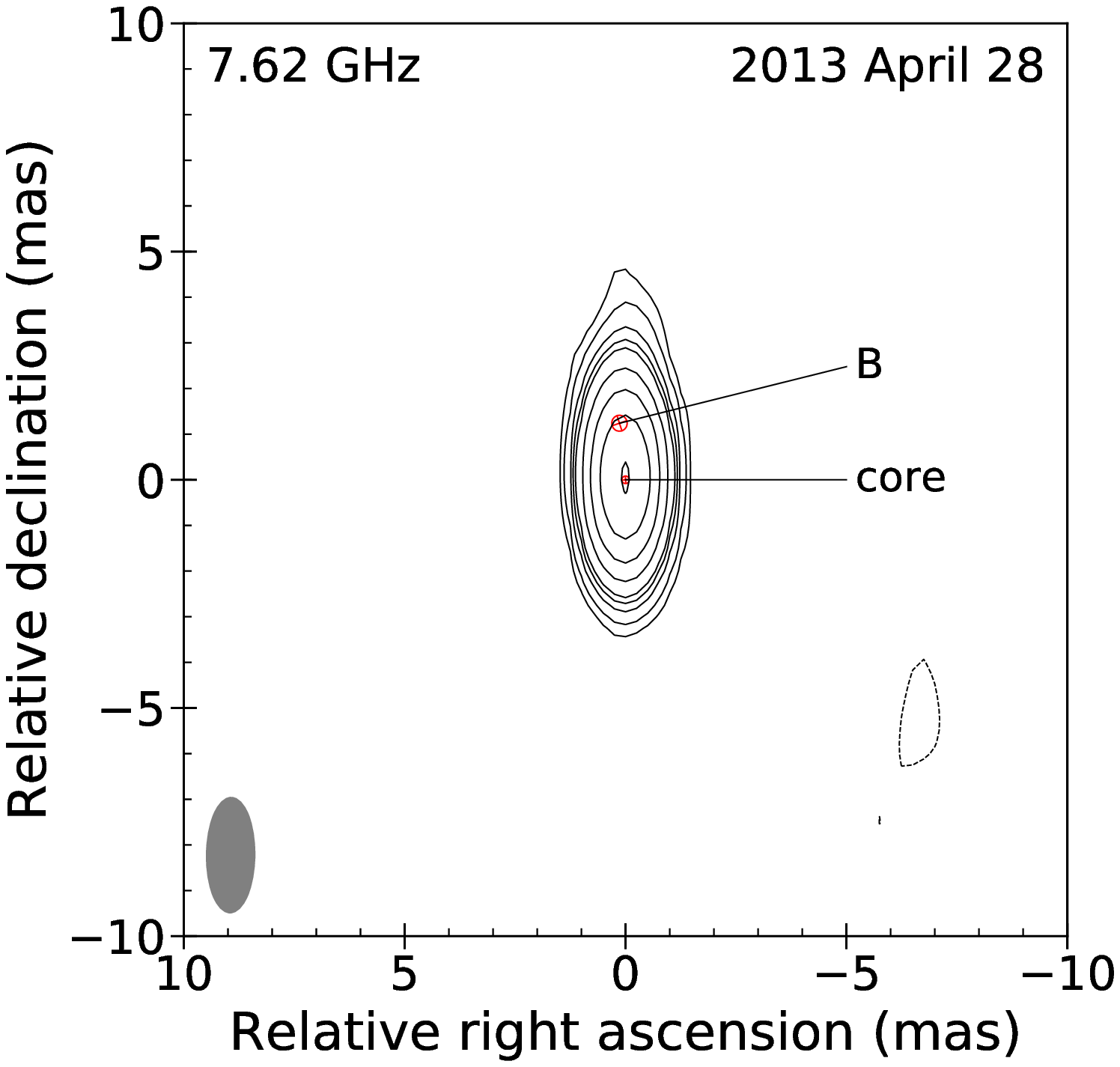}
    \caption{Naturally weighted clean map at 7.6~GHz from the VLBA observations.  The maximum intensity is $70.0$~mJy~beam$^{-1}$, the rms noise is $0.2$~mJy~beam$^{-1}$, the restoring beam is $1.1$~mas$\times2.5$~mas (at $-0.7\degr$ position angle).  The first contours are drawn at $\pm3$ times the rms noise, the positive levels increase by a factor of 2. The fitted model components are denoted with red crossed circles. The Gaussian restoring beam is shown in the bottom left corner.\label{fig:J0909_vlbi2}}
    \label{fig:my_label}
\end{figure}

The structure of the quasar appears less resolved in the 4.3 (Fig.~\ref{fig:J0909_vlbi}a)  and 7.6~GHz VLBA (Fig.~\ref{fig:J0909_vlbi2})  images compared to the global (Fig.~\ref{fig:J0909_vlbi}c) and EVN (Fig.~\ref{fig:J0909_vlbi}b) measurements, due to the larger restoring beam that is  more elongated to the north--south direction. However, the jet direction  towards the north is still discernible. Model fitting to the 4.3-GHz VLBA data set resulted  in 3 components (Table~\ref{tab:J0909_model}); the peak intensity is $101.7\pm4.3$~mJy~beam$^{-1}$, while the integral flux density of the model components is $108.9\pm7.2$~mJy (within $\sim50$~pc projected distance). Our  best-fit model for the 7.6~GHz VLBA data consists of 2 components (Table~\ref{tab:J0909_model}). The integral flux density of the fitted Gaussian model components is $75.1\pm5.5$~mJy (within $\sim10$~pc projected distance), while the peak intensity of the \textsf{clean} map is $70.0\pm3.5$~mJy~beam$^{-1}$.

Using the fitted model parameters (Table~\ref{tab:J0909_model}), we calculated the apparent brightness temperatures ($T_\mathrm{b}$) for the core components at each VLBI epoch, applying the
\begin{equation}
    T_\mathrm{b}=1.22 \times 10^{12} \, (1+z) \, \frac{S}{\vartheta^2\nu^2}\,\mathrm{K}
\end{equation}
formula \citep[e.g.][]{1982ApJ...252..102C}. Here $S$ is the flux density in Jy, $\vartheta$ is the angular size of the fitted circular Gaussian model component (full width at half maximum, FWHM) in mas,  and $\nu$ is the observing frequency in GHz. Values for the global, VLBA and EVN observations are $T_\mathrm{B,G}>0.13\times10^{11}$~K, $T_\mathrm{B,VLBA,4.3}=3.2\times10^{11}$~K, $T_\mathrm{B,VLBA,7.6}=2.1\times10^{11}$~K, and  $T_\mathrm{B,EVN}=2.9\times10^{11}$~K, respectively.

We also calculated Doppler factors according to the
\begin{equation}\label{eq:Tbdoppler}
\delta=\frac{T_\mathrm{b}}{T_\mathrm{b,int}}
\end{equation}
equation, where  $T_\mathrm{b,int}$ is the intrinsic brightness temperature of the source. Lower and upper limits of the Doppler factor were given by applying $T_\mathrm{b,int} \approx 5\times 10^{10}$~K  \citep{1994ApJ...426...51R} and $T_\mathrm{b,int} \approx 3 \times 10^{10}$~K \citep{2006ApJ...642L.115H}. The former corresponds to the equipartition state between the energy densities of the emitting plasma and the magnetic field, while the latter is a characteristic value determined for pc-scale AGN jets based on VLBI observations. The resulting values of the Doppler factor for the VLBA and EVN data sets are between $\delta_\mathrm{eq}=4-6$ and $\delta_\mathrm{char}=7-10$ (Table~\ref{tab:J0909_model}, Cols.~8 and 9). Constrained by the upper limit on the model component FWHM of the global VLBI measurement, only lower limits were found for the values of brightness temperatures and Doppler factors in 1992.

Applying the following equation \citep{1993ApJ...407...65G},
\begin{equation}
\delta_\mathrm{IC}=f(\alpha)S_\mathrm{r}\left(\frac{\ln\frac{\nu_\mathrm{b}}{\nu_\mathrm{r}}}{S_\mathrm{X}\vartheta^{6-4\alpha}\nu_\mathrm{X}^{-\alpha}\nu_\mathrm{r}^{5-3\alpha}} \right)^\frac{1}{4-2\alpha}(1+z),
\end{equation}
a lower limit to the Doppler factor can be calculated, assuming that the X-ray emission originates from the synchrotron self-Compton process of the quasar jet\footnote{This correlation is valid for a discrete jet. For a continuous jet, the $\delta_\mathrm{IC,cont}=\delta_\mathrm{IC}^{(4-2\alpha)/(3-2\alpha)}$ transformation should be applied. In the case of J0909+0354, there is little difference between the values for the two jet models.}, where $S_\mathrm{r}$, $\nu_\mathrm{r}$, and $\vartheta$ are the flux density (Jy), frequency (GHz) and FWHM diameter (mas) of the core model component from the given VLBI observation, respectively, $S_\mathrm{X}$ and $\nu_\mathrm{X}$ are the flux density (Jy) and energy (keV) of the X-ray emission, while $\alpha=-0.75$ is the spectral index (assumed value, e.g. \citealt{1993ApJ...407...65G}), $\nu_\mathrm{b}=10^{5}$~GHz is the cutoff frequency of the high-energy radiation, and $f(\alpha)=-0.08\alpha+0.14$. Values of the inverse-Compton Doppler factor are $\delta_\mathrm{IC}>2$ for both the EVN and VLBA observations, while it is $\delta_\mathrm{IC}>0.1$ for the global VLBI data set, making the two independent estimates of the Doppler factor consistent with each other.

\section{Discussion}
\label{discussion}

\subsection{Inner Jet Structure}\label{ss:inner}

Both the 1992 global VLBI and 2019 EVN observations show a complex pc-scale morphology with a compact core and a jet extending northwards, which is also hinted by the morphology in the $4.3$~GHz VLBA image (Figs.~\ref{fig:J0909_vlbi}c, \ref{fig:J0909_vlbi}a, and \ref{fig:J0909_vlbi}b, respectively). However, as the major axis position angle of the elongated restoring beams in both the global and the VLBA snapshot observations almost perfectly coincides with the jet direction, in the following we will only discuss the  jet structure in details based on the new EVN observations. The mas-scale radio image at 5~GHz (Fig.~\ref{fig:J0909_vlbi}b) reveals a  jet morphology extending up to $\sim20$~mas (150~pc) towards the north, with respect to the core component (i.e. the synchrotron self-absorbed base of the radio jet). 

The jet becomes more diffuse at $\sim10$~mas ($\sim75$~pc), and apparently splits into two branches. The observed division  can be explained by the radio jet interacting with a denser region of the surrounding interstellar medium \citep[e.g.,][]{1999ApJ...518L..87A,2013MNRAS.433..147D}. 
Alternatively, the fork-like  morphology can be attributed to the spine--sheath structure of the jet. The model, supported by numerical simulations as well  \citep[e.g.,][]{2006MNRAS.368.1561M,2007MNRAS.380...51K}, states that the inner region (spine) of the AGN jet propagates with relativistic speeds, while it is surrounded by a slower sub-relativistic sheath \citep[e.g.,][]{1990SvAL...16..284K}. Similar structure can be found in the pc-scale jets of, e.g., Mrk~501 \citep{2004ApJ...600..127G}, 3C~66A (0219+428) and 3C 380 \citep[1828+487; Figs.~2 and 4 in][]{2013AJ....146..120L}, 4C~76.03  \citep[0404+768; Fig.~5 in][]{2013MNRAS.433..147D}, 3C~84 \citep{2014ApJ...785...53N}, 1308+326 \citep{2017A&A...602A..29B}, and S5~0836+710 \citep{2020A&A...641A..40V}.  The spine--sheath structure has recently been reported in one of the best studied cases of the core--jet morphology in the quasar 3C\,273 \citep{2021arXiv210107324B}. Seven out of these eight quasars have complex morphology on kpc scales, i.e.,  hotspots, multiple components, extended jets, or radio lobes \citep{1971ApJ...169....1K,1990MNRAS.246..477P,1991MNRAS.248...86W,1992A&A...266...93H,1993MNRAS.264..298M,1993ApJS...86..365P,1995A&AS..112..235A,1995ApJS...99..297X,1996ApJS..107...37T,1999A&AS..139..601C,2012A&A...545A..65P,2017A&A...601A..35P}, while 4C 76.03 has a compact structure, unresolved with the VLA  \citep{1995ApJS...99..297X}. Similarly to our target source, J0909+0354, a single hotspot was identified in the kpc-scale jet of 3C~273 \citep[e.g.][]{1986Natur.323..419M,2017A&A...601A..35P}.

The structured jet has a pronounced footprint in the linearly polarized emission \citep[e.g.][]{2005MNRAS.356..859P,2013MNRAS.430.1504M}, and it is also visible in the full polarized intensity in the manner of the relative brightening of the outer regions further away from the jet axis  \citep[][]{2004ApJ...600..127G,2005A&A...432..401G,2014ApJ...785...53N,2018NatAs...2..472G,2020A&A...633L...1R}. The influence of external processes (such as the effect of the surrounding medium) on the observed properties of the jet are negligible in the spine--sheath model \citep[e.g.][]{2006MNRAS.368.1561M}. 

A similar approach was discussed in the framework of the two-fluid jet model by \citet{1989A&A...224...24P}, in which  both the superluminal motion at pc scales, and hotspots at kpc scales are explained by an outer thermal electron--proton flow (propagating at non-relativistic speeds, called jet) and an inner relativistic electron--positron plasma (called beam). The apparent split in the pc-scale radio jet (also referred to as limb-brightening) occurs at a distance where the magnetic field becomes weaker than a critical value, hence allowing the relativistic and thermal components to interflow. 
The two-fluid model accounts the observed one-sidedness to the different fraction of the relativistic flow components in the jet and counter-jet (at pc scales), and thus the asymmetric re-acceleration of the thermal flow (resulting in hotspots at kpc scales), rather than the effect of  Doppler-beaming/deboost. The model was recently applied to e.g.  3C\,273 \citep{2021arXiv210107324B}, was proposed to explain \citep{2015IAUS..313..211K} the structure of multiple sources showing both blazar and Fanaroff--Riley type II characteristics \citep[e.g.][]{2006ApJ...637..183L,2010ApJ...710..764K,2015IAUS..313..211K}, and was addressed in the discussion of the blazar PKS\,0735+178 which is reported to show a Fanaroff--Riley type II kpc-structure \citep{2009MNRAS.399.1622G}.

\begin{figure}
    \centering
    \includegraphics[width=\linewidth]{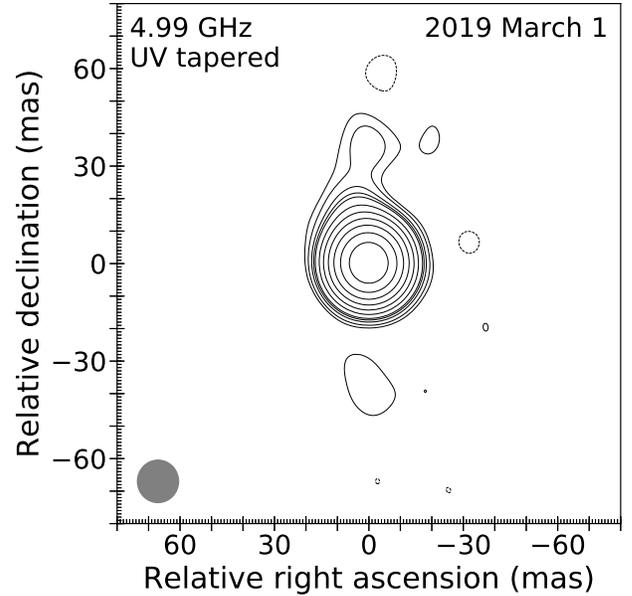}
    \caption{Tapered clean map at 5~GHz from the EVN observations. The maximum intensity is $68.72$~mJy~beam$^{-1}$, the rms noise is $0.07$~mJy~beam$^{-1}$. The first contours are drawn at $\pm3$ times the rms noise, the positive levels increase by a factor of 2. The circular Gaussian restoring beam (13~mas) is shown in the bottom left corner.}
    \label{fig:evntaper}
\end{figure}

To further emphasize the diffuse emission of the pc-scale jet, we applied a Gaussian $(u,v)$ taper (i.e., a scheme where the weights of the visibilities decrease as a function of $(u,v)$ radius) with a value of 0.2 at 10 million wavelength radius to the EVN data set, and repeated the hybrid mapping and Gaussian model fitting procedure. The tapered image (Fig.~\ref{fig:evntaper}) confirms the bending of the jet which is indicated by the weak ($0.35\pm0.08$~mJy flux density) model component found at $34.5\pm2.4$~mas from the core (at $ 2\degr\pm4\degr$ position angle).

The relative positions of VLBI model components (Table~\ref{tab:J0909_model}) and the overall shape of the jet imply that the emission continues towards the NNW component identified in the VLA radio map (as well as in the \textit{Chandra} image), but is resolved out by the EVN between $\sim100$-pc and kpc scales. Moreover, model component positions indicate a bending trajectory: the jet shows a slight turn towards the northwestern direction by $30\degr$ between $65$~pc (C1) and $20$~kpc (NNW).  Such apparent bending have been detected in numerous blazar jets \citep[e.g.,][]{1993ApJ...411...89C,2004A&A...417..887H,2010ApJ...710..764K,2011A&A...529A.113Z,2016ApJ...827...66S,2018MNRAS.477.1065P}. In our case, such morphology might appear interrupted between the components C1 and NNW due to an insufficient brightness sensitivity of the observing system.

We note that the positional discrepancy between the two VLBA data sets,  the absence of the outer jet component at 7.6~GHz, and the large difference between the core flux densities at the two observing frequency bands can be explained as some of the inner jet components at 7.6~GHz are blended into the core at 4.3-GHz, and  that the outer components (i.e. the counterparts of B1 and B2) are too faint to be detected at 7.6~GHz due to the steepening of the spectrum of the jet further away from the core.

\subsection{Jet Parameters}
Brightness temperatures determined from VLBA and EVN measurements well exceed both the theoretical   \citep[$T_\mathrm{b,int} \approx 5\times 10^{10}$~K ,][]{1994ApJ...426...51R} and the somewhat lower empirical \citep[$T_\mathrm{b,int} \approx 3 \times 10^{10}$~K,][]{2006ApJ...642L.115H} limits. The relativistic enhancement is thus clearly indicated by the high values of the Doppler factor ($\delta_\mathrm{eq}=4-6$, $\delta_\mathrm{char}=7-10$). Flux densities at different-epoch VLBI observations reveal the variability of the source at pc scales. We note that  due to the improper resolution of the interferometer (i.e. the upper limit on the FWHM diameter of the core), only lower limits could be determined for the brightness temperature and Doppler factors of the global VLBI observation, therefore we excluded these data from further analysis of the jet parameters.

\begin{figure}
    \centering
    \includegraphics[width=\linewidth]{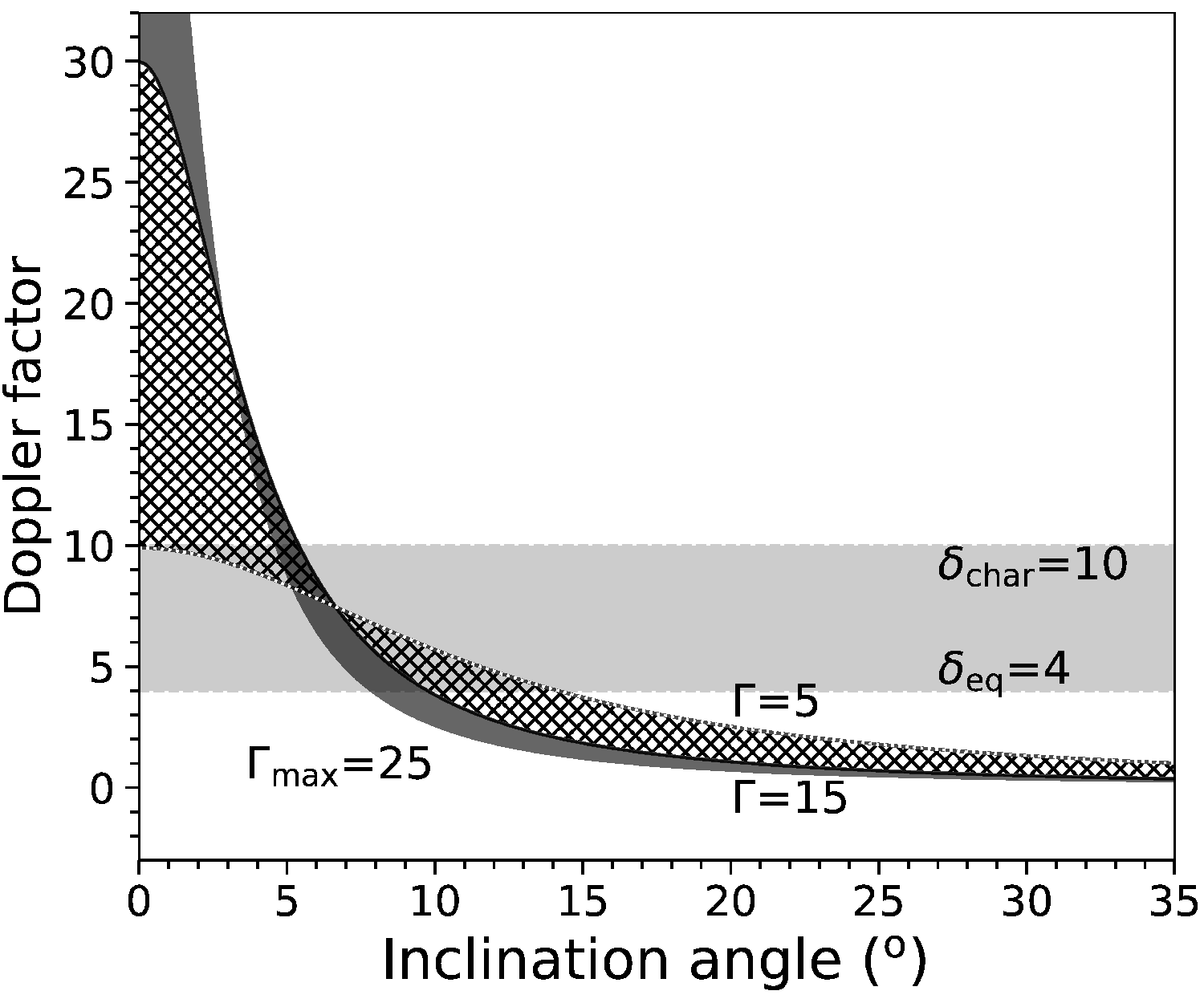}
    \caption{Doppler factors as a function of the inclination angle of the jet with respect to the line of sight to the observer. The gray area denotes the possible Doppler-factor range based on the EVN and VLBA observations. The dashed and dark gray areas denote the allowed parameters for Lorentz factors in the ranges $\Gamma=5-15$ and $\Gamma=15-25$, respectively.}
    \label{fig:doppler}
\end{figure}

Using parameters derived from the models fitted to the visibility data of the EVN observation, we estimated the inclination angle of the jet with respect to the line of sight of the observer. We chose the value of the bulk Lorentz factor to be between $\Gamma=5$ and $\Gamma=15$ \citep[typical values found for high-redshift AGN jets, e.g.][]{2011MNRAS.416..216V}. The estimated inclination angle of the jet is in the range $8\degr\leq\theta_\mathrm{eq}\leq14\degr$ and  $0\degr\leq\theta_\mathrm{char}\leq7\degr$  assuming the values of the equipartition and empirical Doppler factors, respectively. Assuming the empirical upper limit on the bulk Lorentz factor  \citep[$\Gamma=25$, determined in a pc-scale proper motion study of a large sample of AGN jets,][]{2004ApJ...609..539K} for our calculation, the viewing angle of the jet is constrained to the ranges of $6\degr\leq\theta_\mathrm{eq}\leq8\degr$ and  $5\degr\leq\theta_\mathrm{char}\leq6\degr$, for the equipartition and empirical Doppler factors, respectively. The possible ranges of jet parameters are illustrated in Fig~\ref{fig:doppler}. We note that using the inverse-Compton Doppler factor ($\delta_\mathrm{IC}>2$), upper limits on the jet inclination angle of $\theta_\mathrm{IC}\leq23\degr$, $\theta_\mathrm{IC}\leq14\degr$, and $\theta_\mathrm{IC}\leq11\degr$ can be derived by applying bulk Lorentz factors of $\Gamma=5$, $\Gamma=15$, and $\Gamma=25$, respectively.

\subsection{Radio Jet Proper Motion Based on the VLBA and EVN Measurements}

Based on the circular Gaussian model components fitted to the 2013 VLBA and 2019 EVN visibility data, we estimated the apparent proper motion in the pc-scale jet of J0909+0354. Although, including the 1992 global VLBI observations could make the estimation more accurate (due to the longer time span), we do not consider this model as a starting point, because of the unfortunate network geometry.

Since both the 2013 and 2019 data are characterized by a core and two additional jet components, we assumed that the B1 and B2 components in 2013 (VLBA data, 4.3~GHz)  correspond to C1 and C2 in 2019 (EVN data, 5~GHz), respectively. This is further supported by the fact that the component position angles are equal within the uncertainties (Table~\ref{tab:J0909_model}, Col.~5). Over the 5.84-yr time span, the calculated proper motion values for the components B1--C1 and B2--C2 are $\mu_1=0.30\pm0.03$~mas~yr$^{-1}$ and $\mu_2=0.37\pm0.02$~mas~yr$^{-1}$, respectively.  These correspond to  $\beta_1=(31\pm3)~c$ and $\beta_2=(39\pm3)~c$ apparent superluminal speeds, considering the cosmological time dilatation. Although apparent proper motion values as high as our estimates ($\beta_1$ and $\beta_2$) are presented in the literature \citep[e.g. 13  AGN  with $\beta_\mathrm{app}>20$, 3  AGN with $\beta_\mathrm{app}>30$,][]{2004ApJ...609..539K}, both the lowest and highest values of the bulk Lorentz factor determined for either of the component transverse speeds ($\Gamma_\mathrm{min}=53$ and $\Gamma_\mathrm{max}=192$) are much higher than the range of $5\lesssim\Gamma \lesssim 15$ determined for $z>3$ AGN jets \citep{2011MNRAS.416..216V}, and even the lowest value well exceeds the empirical maximum  of $\Gamma=25$  \citep[e.g.][]{2004ApJ...609..539K,2016AJ....152...12L,2017MNRAS.468.4992P}, thus making the determined proper motions rather questionable. We note that due to the slightly different frequency and the different resolutions, the apparent core position (with the core--jet emission blended together) may be different in the VLBA and EVN radio maps, resulting in an apparently different core--jet separation. In any case, new sensitive follow-up VLBI observations to be conducted at 5~GHz in the next 5--10 yr could settle the issue.

\begin{figure*}
\centering
\includegraphics[width=.8\linewidth]{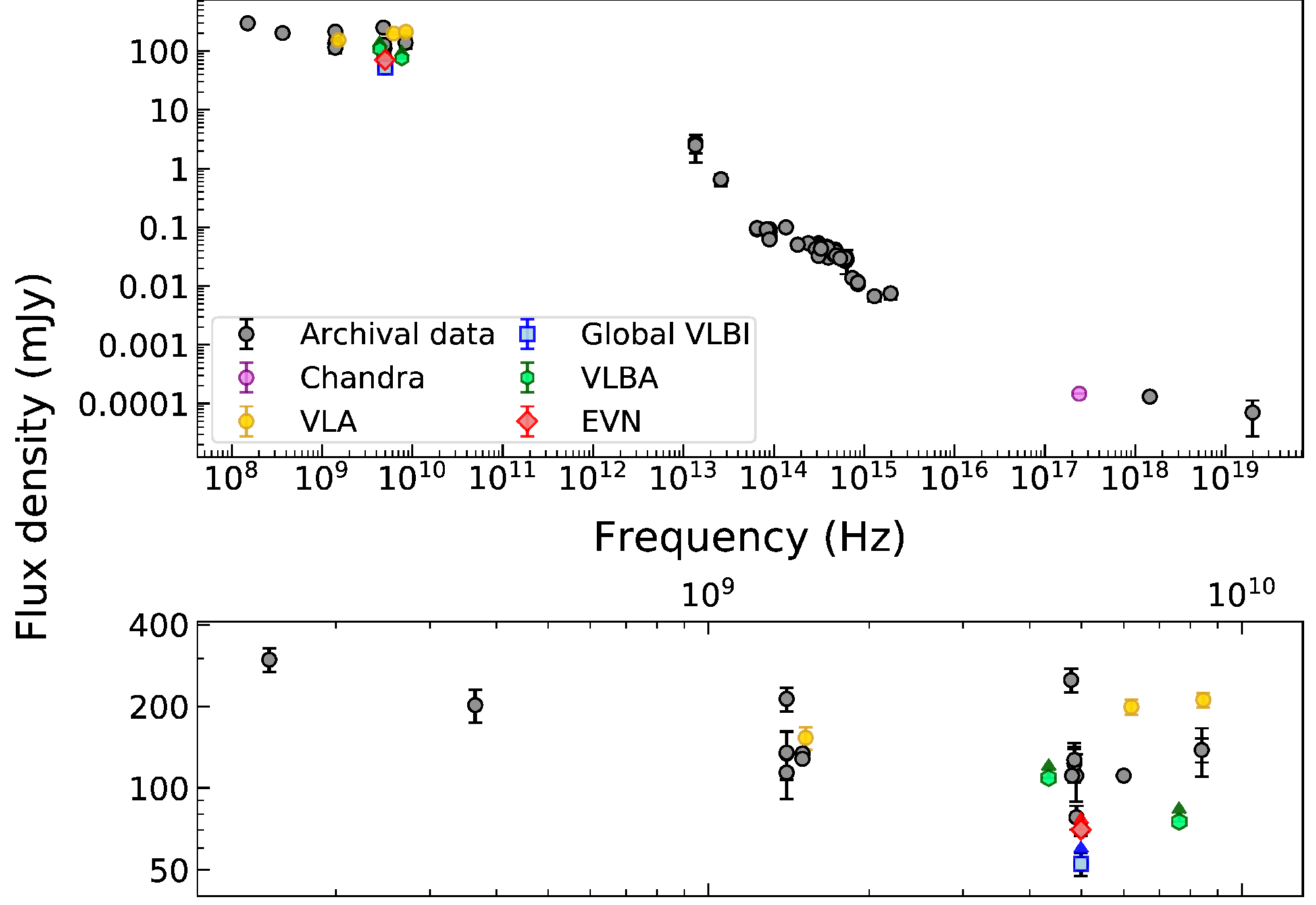}
\caption{Broad-band spectrum of J0909+0354. Black circles denote data from archival radio--X-ray observations and are acquired from the NASA/IPAC Extragalactic Database and the photometry tool of VizieR service. Flux densities from the \textit{Chandra} data set are shown in purple, while blue, green, red and yellow symbols denote the total flux density of the global VLBI, VLBA, EVN and VLA measurements, respectively. For the better visibility of the radio part of the spectrum, these data are also shown in the lower panel.}\label{fig:J0909_sed}
\end{figure*}

\subsection{Broad-band and Radio Spectra}

The broad-band and radio spectra of the quasar are shown in the upper and lower panels of Fig.~\ref{fig:J0909_sed}, respectively. Data were obtained from the NASA/IPAC Extragalactic Database\footnote{\url{https://ned.ipac.caltech.edu/}}, the photometry tool of the VizieR service\footnote{\url{http://vizier.unistra.fr/vizier/sed/}}, and from  \textit{Chandra}, VLA, global VLBI, VLBA, and EVN observations discussed in this paper. 

The differences in total flux densities at similar frequencies but different epochs can partly be attributed to the different angular resolutions of the iffnterferometer arrays \citep[FIRST, NVSS, and CLASS,][]{2015ApJ...801...26H,1998AJ....115.1693C,2003MNRAS.341....1M} and single-dish \citep[PMN, GBT,][]{1995ApJS...97..347G,1991ApJS...75....1B,1992ApJS...79..331W,1996ApJS..103..427G} observations. 
We fitted  the  logarithm of the flux density values of the entire radio waveband (from 150~MHz to $8.4$~GHz) with a linear curve, and found a spectral index of  $\alpha_\mathrm{kpc}=-0.13\pm0.06$ for the kpc-scale structure. 

The flux density variability of the pc-scale emission is $35-50\%$ between the epochs of the global VLBI, VLBA, and EVN observations at $\sim 5$~GHz.
We calculated radio spectral indices for the pc-scale structure using the flux densities determined from the simultaneous dual-frequency VLBA observations and the new EVN data. The power-law spectral index in the  $4.34-7.62$~GHz frequency range is  $\alpha_{4.34}^{7.62}=-0.66\pm0.17$ and  $\alpha_\mathrm{pc}=-0.46\pm0.13$, for the VLBA and VLBA--EVN total flux density data, respectively. Similarly, spectral indices for the core components are  $\alpha_{4.34,\mathrm{core}}^{7.62}=-0.74\pm0.22$ and $\alpha_\mathrm{pc,core}=-0.55\pm0.15$.  On the one hand, as the quasar shows significant flux density variability on pc scales (Table~\ref{tab:J0909_model}), these values should be treated with caution. Although parallel observations with the VLBA are not affected by variability, the steep  spectrum can be the result of the different angular resolution at the two frequencies, thus the comparisons of the components and flux densities remain uncertain.  On the other hand, steep spectra with $-0.52\leq\alpha_\mathrm{core}\leq-1$ were previously found for high-redshift ($z\geq4$) quasars, albeit with not Doppler-boosted radio emission \citep[e.g.][]{2003MNRAS.343L..20F,2005A&A...436L..13F,2008A&A...484L..39F,2010A&A...524A..83F,2011A&A...531L...5F,2016MNRAS.463.3260C,2017MNRAS.467..950C}, the blazar PSO~0309+27 \citep[at $z=6.1$,][]{2020A&A...643L..12S} between $1.5$~GHz and $5$~GHz frequencies, as well as    J0906+6930 ($z=5.47$) in the $15$~GHz$\leq\nu\leq43$~GHz frequency range \citep[although its core spectral index flattens to $\alpha_\mathrm{core}=0.2$ below 8.4~GHz,][]{2017MNRAS.468...69Z}.

\subsection{Kpc-scale Structure: the NNW Component}
\label{ss:nnw}

As noted in subsection \ref{ss:kpc-str}, the radio spectrum of the NNW component between 1.5~GHz and 8.5~GHz is consistent with its identification as a jet hotspot. Not that many high-redshift sources demonstrate well-pronounced kpc-scale jets and/or extended morphological features which might be suspected as being jets. An increasing rareness of detectable jets at both pc and kpc scales in AGN with increasing redshift is rather natural \citep[e.g.,][]{Gurvits-SKA-1999,2015IAUS..313..327G}. Nevertheless, such examples are known: J1405+0415 at $z=3.215$ \citep{1992A&A...260...82G,2008A&A...489..517Y}, J2217+0220 at $z=3.572$ \citep{2001ApJ...547..714L}, and J2134$-$0419 \citep{2018MNRAS.477.1065P} at $z=4.33$, to list just three of them.

\begin{figure*}
\centering
\includegraphics[width=0.45\linewidth]{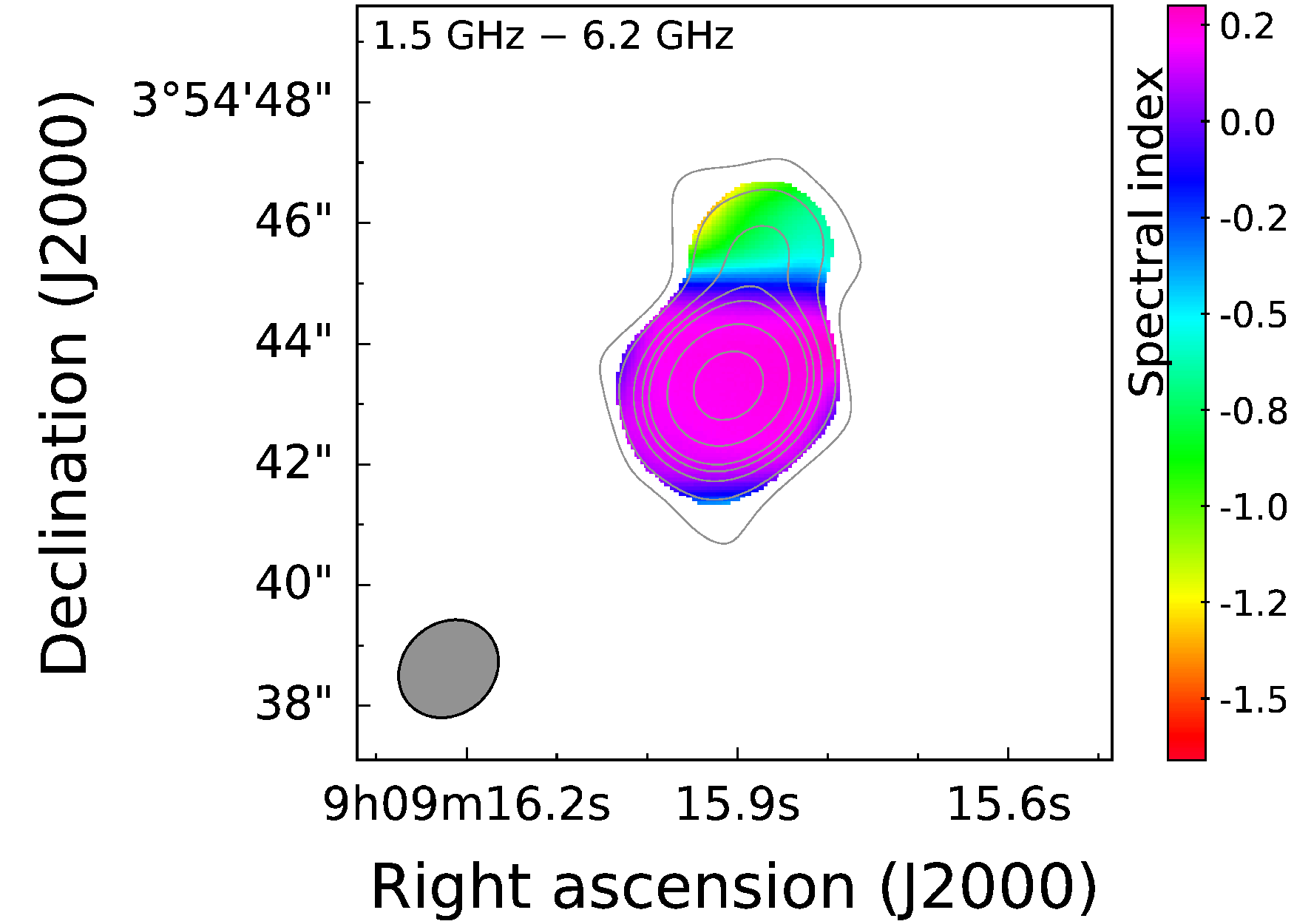}
\includegraphics[width=0.45\linewidth]{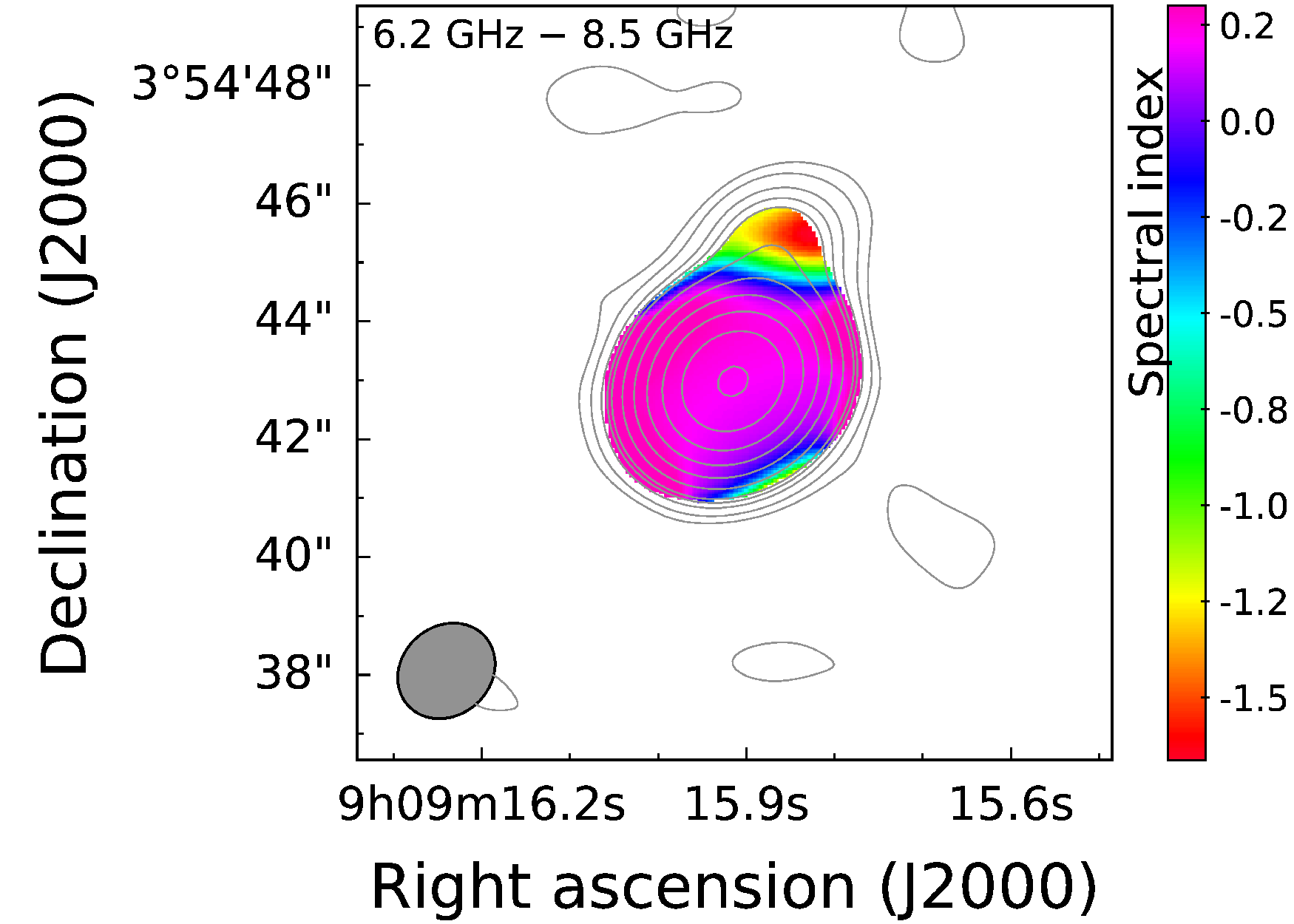}
\caption{Spectral index maps of the quasar J0909$+$0354 between 1.5 and 6.2~GHz (left) and 6.2 and 8.5~GHz (right). We applied a Gaussian $(u,v)$ taper with a value of 0.2 at $10^{5}$ wavelength radius on the 6.2 and 8.5~GHz data. The clean maps were created using identical gridding ($0\farcs05$ per pixel) and restoring beam ($1\farcs51\times1\farcs75$ at$-49\degr$ position angle). Contour lines denote the clean models of the 1.5~GHz (left) and 6.2~GHz data (right), respectively, starting at $\pm3$ times the rms values ($1.06$~mJy~beam$^{-1}$ and $0.12$~mJy~beam$^{-1}$, respectively), with the levels increasing by a factor of 2.} \label{fig:alphamap}
\end{figure*}

Two-point spectral indices  ($\alpha_\mathrm{6.2NNW}^\mathrm{1.5}=-0.94\pm0.19$, $\alpha_\mathrm{7.77,NNW}^\mathrm{4.62}=-1.03\pm 0.10$, $\alpha_\mathrm{8.5,NNW}^\mathrm{6.2}=-2.05\pm0.25$)  calculated from  the VLA observations for the NNW component show a spectral steepening with increasing frequency, which is also indicated in the $1.5-6.2$~GHz and $6.2-8.5$~GHz spectral index maps in Fig.~\ref{fig:alphamap}. Interpreting NNW as the approaching hotspot of the quasar, we expect it to have a flatter spectrum than the receding one \citep[e.g.][]{1997MNRAS.289..753D,2000MNRAS.317..658I}. Thus non-detection of the other (receding) hotspot can be explained by the fact that the spectral index difference is enhanced resulting from the small inclination angle of J0909+0354 \citep[e.g.][]{1997MNRAS.289..753D,2000MNRAS.317..658I}. We can apply the formula 
\begin{equation}
 K=\left(\frac{1+\beta\cos\theta}{1-\beta\cos\theta}   \right)^{2-\alpha},
\end{equation}
where $K$ is the ratio of the flux densities of the approaching and receding sides of the jet, $\beta=vc^{-1}$ is the speed of the jet, $\theta$ is  the  jet inclination angle with respect to the line of sight of the observer, and $\alpha$ is the spectral index of the hotspot. With  $\beta=0.3$ \citep[e.g. from][]{1997MNRAS.289..753D}, $\theta=23\degr$ (upper limit calculated for J0909+0354), and single-epoch NNW spectral index $\alpha_\mathrm{7.77,NNW}^\mathrm{4.62}=-1.03$, the  flux density ratio  for the approaching (NNW) and receding hotspots of the kpc-scale jet is $K=5.42$. Flux densities for the hotspot on the receding side are then expected to be approximately $S_\mathrm{1.5,r}=4$~mJy, $S_\mathrm{6.2,r}=1$~mJy, and $S_\mathrm{8.5,r}=0.5$~mJy, for the 1.5, 6.2, and 8.5~GHz  frequency bands, respectively. These values are in the same order of magnitude as the $3\sigma$ rms noise of the VLA clean maps (Fig.~\ref{fig:vla}), which can naturally explain the non-detection of the hotspot on the receding side of the kpc-jet by the array. Considering higher jet speeds or lower jet viewing angles results in even lower flux density estimates. The spectral steepening of NNW with increasing frequency implies radiation losses due to spectral aging \citep[e.g.][]{1991AJ....102.1659K,1999AJ....117..677B,2000MNRAS.317..658I,2019MNRAS.484..385V}.

There is an apparent contradiction between identifying NNW as a hotspot and the proposed spine-sheath feature observed on pc scales (subsection \ref{ss:inner}). This can be resolved by taking into account the different timescales of the pc- and kpc-scale structures. Although the injection to hotspots is expected to originate from a continuous supply of particles, there is a large physical distance between the inner jet ($\sim10$~pc) and the kpc-scale NNW component ($\sim17$~kpc). The $17$~kpc projected distance of NNW with respect to the core component translates to  $43.5$~kpc length considering the upper limit of $\theta_\mathrm{max}=23\degr$ on the jet inclination angle. The presently observed hotspot-like characteristics are not expected to be affected by the more recent state of the pc-scale structure. Assuming the speed of the jet fueling NNW is $0.3\,c$, it takes $\sim0.5$~Myr in the rest frame of the quasar for the newly developed changes in the pc-scale jet to propagate to the kpc-scale structure. The hotspot scenario is further supported by the X-ray detection of enhanced emission at the NNW region, and a jet-like feature connecting it to the core (Section \ref{ss:kpc-str}). Concluding the discussion above, identification of NNW as a hotspot cannot be excluded.

\begin{figure*}
    \centering
    \includegraphics[width=0.45\linewidth]{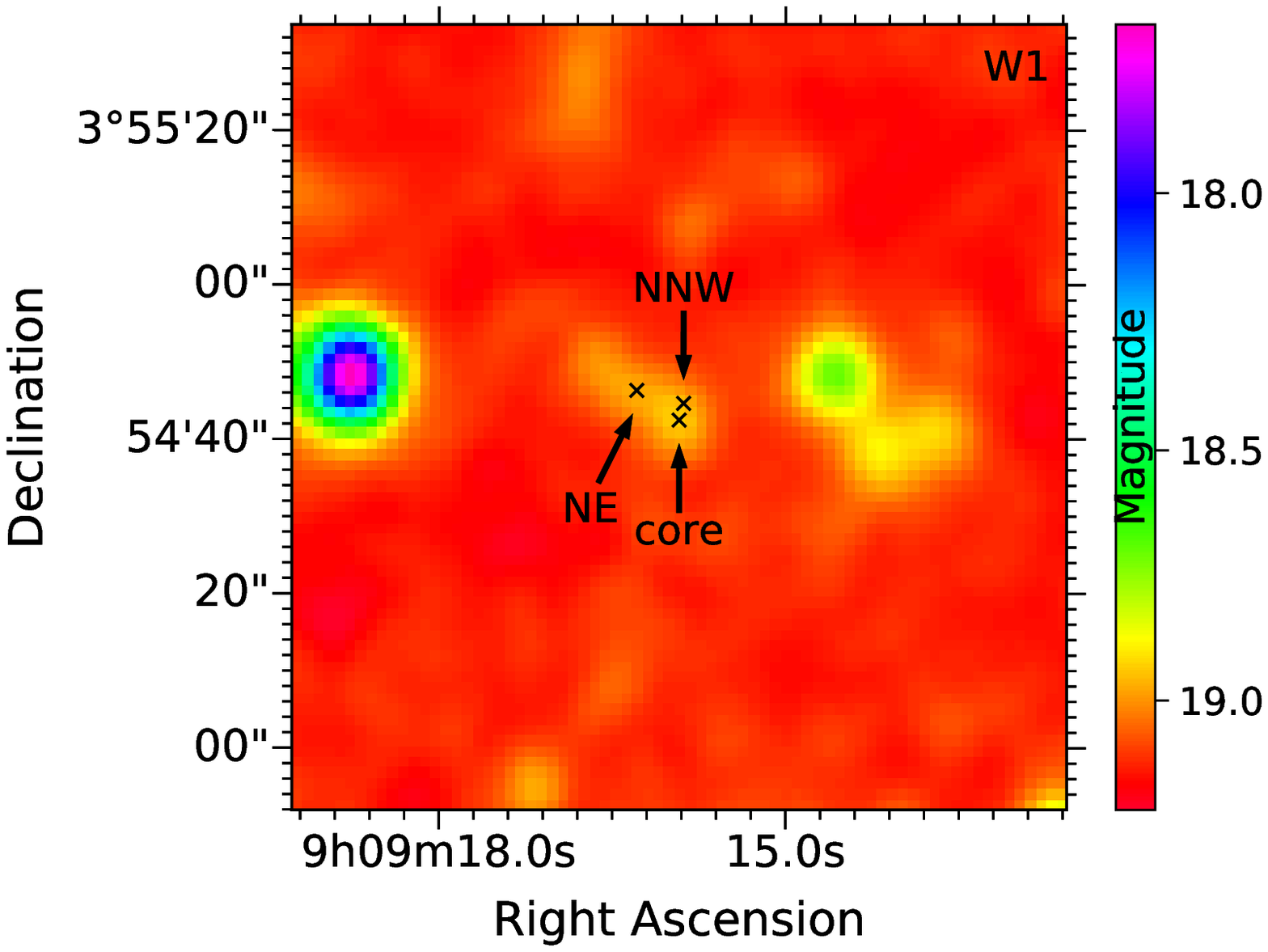} \includegraphics[width=0.45\linewidth]{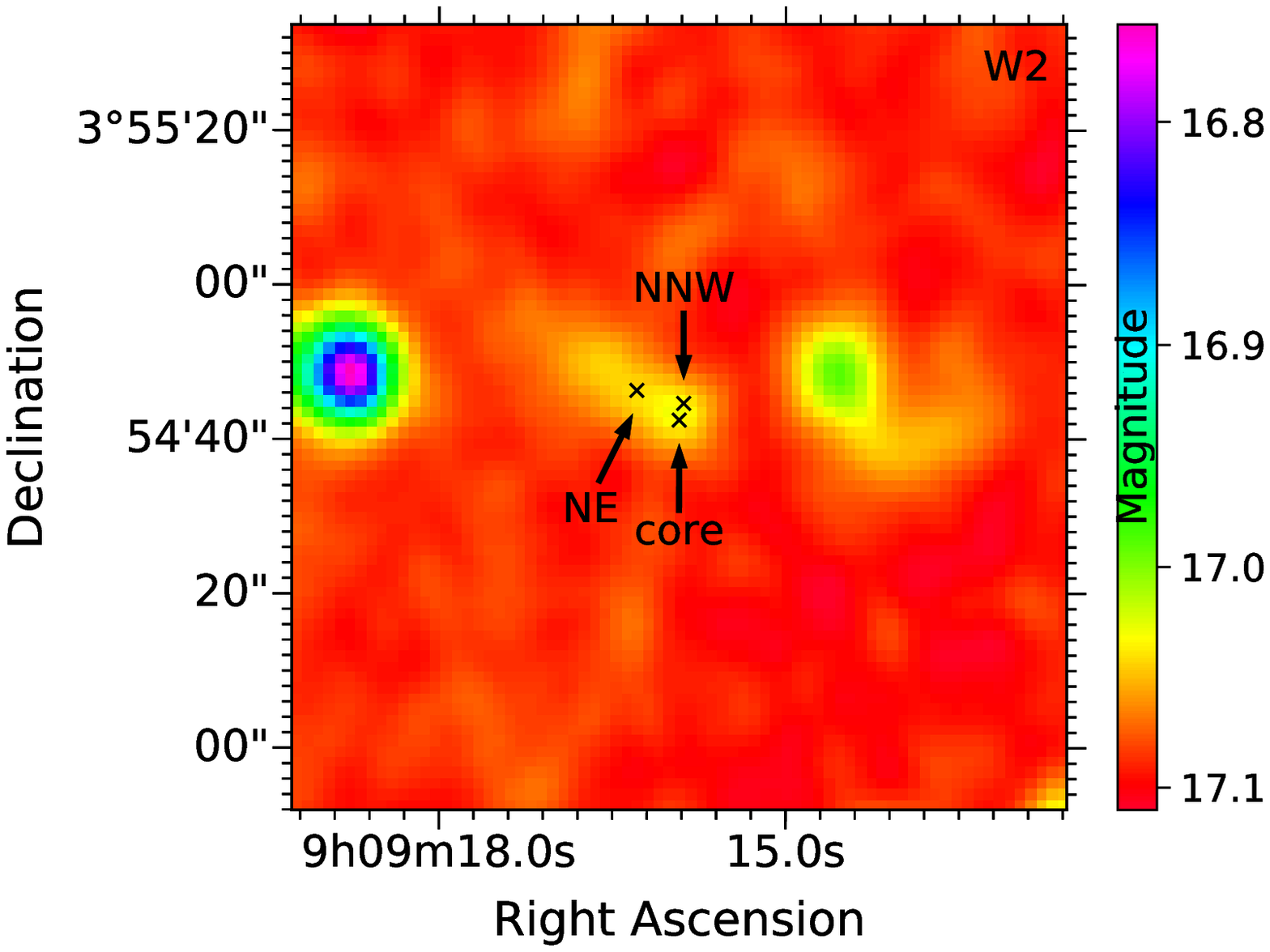}
    \caption{Mid-infrared maps of the quasar and its neighborhood observed with \textit{WISE} in $W1$ and $W2$ bands \citep[AllWISE,][]{2010AJ....140.1868W,2011ApJ...731...53M}. Black crosses denote the positions of the X-ray components detected with \textit{Chandra}. The images were obtained from the \textit{WISE} Science Data Center (\url{https://irsa.ipac.caltech.edu}).}
    \label{fig:wise}
\end{figure*}

\subsection{Kpc-scale Structure: the NE Component}

From the survey of \citet{2016ApJ...819...62C}, the post facto probability of such an unrelated strong source being within 7\arcsec of our target is $0.03\%$. However, with more than 300,000 distinct X-ray sources in the second \textit{Chandra} source catalog \citep{2020AAS...23515405E}, such a probability is not evidence of association with the quasar J0909+0354. Indeed, there is a faint source in the Pan-STARRS images coincident with the NE X-ray source. We estimate $g, r, i, z, y$ magnitudes of $23.89, 22.74, 22.73, 22.63, 21.50$, respectively, from the prescription of \citet{2020ApJS..251....4W}. An optical object is also detected at the position of the NE feature in the $g,r,z$ bands in the Dark Energy Camera Legacy Survey\footnote{Legacy Surveys / D. Lang (Perimeter Institute), \url{https://www.legacysurvey.org/viewer?ra=137.31786883&dec=3.91286574&layer=ls-dr9&zoom=15}} \citep[DECaLS,][]{2019AJ....157..168D}. We note that, contrary to NE, the NNW component has no optical counterpart. With no radio emission at or in the direction of the NE feature, there is no evidence to associate it with J0909+0354, and we will assume here that it is a foreground or background object. 

Mid-infrared \textit{WISE} maps \citep[from the AllWISE data release,][]{2010AJ....140.1868W,2011ApJ...731...53M} centered at the position of J0909+0354 at $3.4~\mu$m ($W1$) and $4.6~\mu$m ($W2$) show an extended emission surrounding the core component and NE \citep[Fig.~\ref{fig:wise}; DN-to-magnitude conversion was carried out as described by][]{2011wise.rept....1C}. The emission can be traced up to $15\arcsec-20\arcsec$ ($\sim130-160$~kpc projected linear distance) with respect to the position of the core. We note that the extended emission can be a blend of  individual sources located at different cosmological distances but seen in projection, considering the angular resolution of \textit{WISE} ($6\farcs1$ and $6\farcs4$ for $W1$ and $W2$, respectively). But a physical connection between the two objects (J0909+0354 and NE) cannot be ruled out based on the data available.

\section{Summary and conclusions}
\label{summary}

Using data from multi-epoch VLBI imaging experiments, we characterized the pc-scale structure of the high-redshift quasar J0909$+$0354. Fitting circular Gaussian model components to the visibility data of global VLBI, VLBA, and EVN measurements, we found a Doppler-enhanced core and multiple jet components. The inner jet is extended towards the north, i.e. appears to be related to the NNW component of the kpc-scale radio structure seen in the 1.5, 6.2, and 8.5-GHz VLA images, as well as the X-ray jet in the \textit{Chandra} image. We discussed the possible nature of the NNW  component, using VLA observations at 1.5, 6.2, and 8.5~GHz. Although its possible identification as a hotspot is supported by its steep radio spectrum and the X-ray detection with \textit{Chandra}, the one-sidedness of the kpc-scale extended structure may challenge this interpretation. The high brightness temperatures of the core components (and hence the high Doppler factors), the estimated small viewing angles with respect to the line of sight to the observer, the $\sim30\degr$ bending of the jet between pc and kpc scales, and the flux density variability of the quasar are all characteristics of a blazar-type AGN. Measurements of the apparent proper motion of pc-scale jet components and determination of the jet inclination angle and bulk Lorentz factor proved to be difficult because of the unfavorable restoring beam orientation in the first-epoch VLBI experiment in 1992. Future 5-GHz VLBI observations could provide sufficient data for refining our estimates of the inner jet inclination ($\theta \lesssim 8\degr$). The apparent jet bending between pc and kpc scales in J0909$+$0354 indicated also by our tapered EVN image could possibly be directly observable with medium-resolution ($\sim 100$~mas) radio interferometric imaging.

Based on data from archival observations and the $6.2$~GHz VLA observation, we studied the radio spectrum of the quasar at kpc scales, which resulted in an overall radio spectral index of $\alpha_\mathrm{kpc}=-0.13\pm0.06$.  We also determined the pc-scale spectral indices, based on the fitted model parameters to the core component of the the dual-frequency VLBA and the new EVN observations. We found the values $\alpha_{4.34,\mathrm{core}}^{7.62}=-0.74\pm0.22$ and $\alpha_\mathrm{pc,core}=-0.55\pm0.15$ for the two- and three-point spectral indices, respectively. As the three-point spectral index $\alpha_\mathrm{pc,core}$ could be considered flat (i.e. $\gtrsim -0.5$) within the uncertainties, we conclude that the apparent spectral steepness suggested by $\alpha_{4.34,\mathrm{core}}^{7.62}$ can be attributed to the blending of flux densities of different jet components in the dual-frequency VLBA observations. Flux density variability of the source on pc scale can also play a role.

We investigated the additional X-ray component (NE) without a radio counterpart identified in the \textit{Chandra} image located at $\sim 6\farcs5$ separation from the quasar in the northeastern direction. A faint optical counterpart was found for the NE component in Pan-STARRS and DECaLS. We assume that this feature has no physical connection with J0909+0354. It is most likely a foreground or background object seen close to the quasar only in projection. However, the faint elongated (up to $\sim160$~pc projected linear size) mid-infrared emission region containing both the core and NE in the $3.4~\mu$m and $4.6~\mu$m \textit{WISE} images might suggest a physical interaction between J0909$+$0354 and another nearby X-ray and optical quasar located at the NE position. Information about the redshift of the NE source would be needed to unambiguously decide whether the two object are physically close to each other.

\begin{acknowledgements}

The authors thank the anonymous referee for their insightful comments and suggestions.
KP, SF, and K\'EG thank the Hungarian National Research, Development and Innovation Office (OTKA K134213) for support.  DAS is supported by  NASA contract NAS8-03060 to SAO, and grant
GO8-19077X from the CXC. The EVN is a joint facility of independent European, African, Asian and North American radio astronomy institutes. Scientific results from data presented in this publication are derived from the following EVN project code: EP115. The research leading to these results has received funding from the European Commission Horizon 2020 Research and Innovation Programme under grant agreement No. 730562 (RadioNet). The NRAO is a facility of the National Science Foundation operated under cooperative agreement by Associated Universities, Inc. This publication made use of the Astrogeo VLBI FITS image database (http://astrogeo.org/vlbi$\_$images/) and the authors thank Leonid Petrov for making results of the J0909+0354 observations online prior to publication. This research has made use of the NASA/IPAC Extragalactic Database (NED) which is operated by the Jet Propulsion Laboratory, California Institute of Technology, under contract with the National Aeronautics and Space Administration. This research has made use of the VizieR catalogue access tool, CDS, Strasbourg, France (DOI: 10.26093/cds/vizier). The original description  of the VizieR service was published in 2000, A\&AS 143, 23. The Pan-STARRS1 Surveys (PS1) and the PS1 public science archive have been made possible through contributions by the Institute for Astronomy, the University of Hawaii, the Pan-STARRS Project Office, the Max-Planck Society and its participating institutes, the Max Planck Institute for Astronomy, Heidelberg and the Max Planck Institute for Extraterrestrial Physics, Garching, The Johns Hopkins University, Durham University, the University of Edinburgh, the Queen's University Belfast, the Harvard-Smithsonian Center for Astrophysics, the Las Cumbres Observatory Global Telescope Network Incorporated, the National Central University of Taiwan, the Space Telescope Science Institute, the National Aeronautics and Space Administration under Grant No. NNX08AR22G issued through the Planetary Science Division of the NASA Science Mission Directorate, the National Science Foundation Grant No. AST-1238877, the University of Maryland, E\"otv\"os Lor\'and University (ELTE), the Los Alamos National Laboratory, and the Gordon and Betty Moore Foundation. This publication makes use of data products from the Wide-field Infrared Survey Explorer, which is a joint project of the University of California, Los Angeles, and the Jet Propulsion Laboratory/California Institute of Technology, funded by the National Aeronautics and Space Administration.
\end{acknowledgements}

\software {AIPS} \citep{1995ASPC...82..227D,2003ASSL..285..109G}, CASA \citep{2007ASPC..376..127M}, Difmap \citep{1997ASPC..125...77S}, CIAO 4.12  \citep{2006SPIE.6270E..1VF}, Sherpa \citep{2007ASPC..376..543D}, DS9 \citep{2005ASPC..347..110J}, Astropy \citep{2013A&A...558A..33A}, Maptlotlib \citep{2007CSE.....9...90H}

\bibliography{j0909_bib.bib}{}
\bibliographystyle{aasjournal}

\end{document}